\documentclass[12pt]{article}
\usepackage{graphicx}
\def\beq{\begin{equation}}
\def\eeq{\end{equation}}
\def\bea{\begin{eqnarray}}
\def\eea{\end{eqnarray}}
\def\nn{\nonumber}
\def\roughly#1{\mathrel{\raise.3ex\hbox
{$#1$\kern-.75em\lower1ex\hbox{$\sim$}}}}

\def\gsim{\roughly>}
\def\sss{\scriptscriptstyle}
\def\Bbar{{\overline B}^0}
\def\bd{B_d^0}
\def\bdbar{{\overline{B_d^0}}}
\def\bs{B_s^0}
\def\bsbar{{\overline{B_s^0}}}
\def\ks{K_{\sss S}}

\def\ANPq{{\cal A}^q}
\def\ANPd{{\cal A}^d}
\def\ANPu{{\cal A}^u}
\def\ANPs{{\cal A}^s}
\def\ANPc{{\cal A}^c}
\def\ANPqlam{{\cal A}_\lambda^q}

\def\ANPdlam{{\cal A}_\lambda^d}
\def\ANPclam{{\cal A}_\lambda^c}
\def\ANPslam{{\cal A}_\lambda^s}
\def\ANPqi{{\cal A}_i^q}
\def\ANPqo{{\cal A}_0^q}
\def\ANPqperp{{\cal A}_\perp^q}
\def\ANPqpar{{\cal A}_\|^q}
\def\ssbar{(s{\bar s})}
\def\btod{{\bar b} \to {\bar d}}
\def\btos{{\bar b} \to {\bar s}}
\def\bvv{B \to V_1 V_2}
\def\bra#1{\left\langle #1\right|}
\def\ket#1{\left| #1\right\rangle}
\def\barpk{{\raise.35ex\hbox
{${\sss (}$}}--{\raise.35ex\hbox{${\sss )}$}}}
\def\bbarp{\hbox{$B$\kern-0.9em\raise1.4ex\hbox{\barpk}}}

\def\pew{P_{\sss EW}}
\def\pewp{P'_{\sss EW}}
\def\pewc{P_{\sss EW}^{\sss C}}
\def\pewcp{P_{\sss EW}^{\prime\sss C}}
\def\ApNPqph{{\cal A}^{\prime,q} e^{i \Phi'_q}}
\def\ApNPuph{{\cal A}^{\prime,u} e^{i \Phi'_u}}
\def\ApNPdph{{\cal A}^{\prime,d} e^{i \Phi'_d}}
\def\ApNPCqph{{\cal A}^{\prime {\sss C}, q} e^{i \Phi_q^{\prime C}}}
\def\ApNPCuph{{\cal A}^{\prime {\sss C}, u} e^{i \Phi_u^{\prime C}}}
\def\ApNPCdph{{\cal A}^{\prime {\sss C}, d} e^{i \Phi_d^{\prime C}}}
\def\ApNPcomb{{\cal A}^{\prime, comb} e^{i \Phi'}}
\def\ApNPCph{{\cal A}^{\prime {\sss C}} e^{i \Phi^{\prime \sss C}}}
\def\ANPuph{{\cal A}^u e^{i \Phi_u}}
\def\ANPdph{{\cal A}^d e^{i \Phi_d}}

\def\ANPCuph{{\cal A}^{{\sss C}, u} e^{i \Phi_u^{\sss C}}}
\def\ANPCdph{{\cal A}^{{\sss C}, d} e^{i \Phi_d^{\sss C}}}
\def\ANPCph{{\cal A}^{\sss C} e^{i \Phi^{\sss NP,C}}}
\def\ANPcomb{{\cal A}^{comb} e^{i \Phi}}

\def\aI{a_{\sss I}}
\def\aR{a_{\sss R}}

\def\Aut{{\cal A}_{ut}}
\def\Act{{\cal A}_{ct}}
\def\Autp{{\cal A}'_{ut}}
\def\Actp{{{\cal A}'_{ct}}}

\def\lft{{\sss L}}

\def\epjc#1#2#3{{ Eur.\ Phys.\ J.}\ {\bf C#1}, #3 (#2)}

\def\npb#1#2#3{{ Nucl.\ Phys.} {\bf B#1}, #3 (#2)}
\def\plb#1#2#3{{ Phys.\ Lett.} {\bf #1B}, #3 (#2)}
\def\prd#1#2#3{{ Phys.\ Rev.} {\bf D#1}, #3 (#2)}
\def\newprd#1#2#3{{ Phys.\ Rev.} {\bf D#1}, #3 (#2)}
\def\prl#1#2#3{{ Phys.\ Rev.\ Lett.} {\bf #1}, #3 (#2)}

\begin{document}

\begin{flushright}  
UdeM-GPP-TH-04-123 \\ 
McGill 13/04 \\ 
IMSc/2004/06/25 \\
\end{flushright}

\begin{center}
\bigskip
{\Large \bf 
Methods for Measuring New-Physics Parameters \\ \vskip2truemm
in {\boldmath $B$} Decays} \\
\bigskip
{\large 
Alakabha Datta $^{a,b,}$\footnote{datta@physics.utoronto.ca},
Maxime Imbeault $^{b,}$\footnote{maxime.imbeault@umontreal.ca},
David London $^{b,c,}$\footnote{london@lps.umontreal.ca},
V\'eronique Pag\'e $^{b,}$\footnote{veronique.page@umontreal.ca},
Nita Sinha $^{d,}$\footnote{nita@imsc.res.in} and
Rahul Sinha $^{d,}$\footnote{sinha@imsc.res.in}}
\end{center}

\begin{flushleft}
~~~~~~~~~~~$a$: {\it Department of Physics, University of Toronto,}\\
~~~~~~~~~~~~~~~{\it 60 St.\ George Street, Toronto, ON, Canada M5S 1A7}\\
~~~~~~~~~~~$b$: {\it Laboratoire Ren\'e J.-A. L\'evesque, 
Universit\'e de Montr\'eal,}\\
~~~~~~~~~~~~~~~{\it C.P. 6128, succ. centre-ville, Montr\'eal, QC,
Canada H3C 3J7} \\
~~~~~~~~~~~$c$: {\it Physics Department, McGill University,}\\
~~~~~~~~~~~~~~~{\it 3600 University St., Montr\'eal QC, Canada H3A 2T8}\\
~~~~~~~~~~~$d$: {\it Institute of Mathematical Sciences, C. I. T Campus,}\\
~~~~~~~~~~~~~~~{\it Taramani, Chennai 600 113, India}
\end{flushleft}
\newpage

\begin{center} 
\bigskip (\today)
\vskip0.5cm
{\Large Abstract\\}
\vskip3truemm
\parbox[t]{\textwidth} {Recently, it was argued that new-physics (NP)
effects in $B$ decays can be approximately parametrized in terms of a
few quantities. As a result, CP violation in the $B$ system allows one
not only to detect the presence of new physics (NP), but also to
measure its parameters. This will allow a partial identification of
the NP, before its production at high-energy colliders. In this paper,
we examine three methods for measuring NP parameters. The first uses a
technique involving both $\btos$ and $\btod$ penguin $B$ decays.
Depending on which pair of decays is used, the theoretical error is in
the range 5--15\%. The second involves a comparison of $B\to \pi K$
and $B\to\pi\pi$ decays. Although the theoretical error is large
($\gsim 25\%$), the method can be performed now, with
presently-available data. The third is via a time-dependent angular
analysis of $\bvv$ decays. In this case, there is no theoretical
error, but the technique is experimentally challenging, and the method
applies only to those NP models whose weak phase is universal to all
NP operators. A reliable identification of the NP will involve the
measurement of the NP parameters in many different ways, and with as
many $B$ decay modes as possible, so that it will be important to use
all of these methods.}
\end{center}

\thispagestyle{empty}
\newpage
\setcounter{page}{1}
% Decrease texheight (for preprint numbers) again
%\textheight 23.0 true cm
\baselineskip=14pt

\section{Introduction}

Within the standard model (SM), CP violation is due to the presence of
a complex phase in the Cabibbo-Kobayashi-Maskawa (CKM) matrix
\cite{pdg}. The principal goal of the study of CP violation in the $B$
system is to test this explanation, and to find evidence for physics
beyond the SM. As a result, much theoretical work has concentrated on
signals of new physics (NP) in $B$ decays \cite{CPreview}.

At present, we have several experimental hints of new physics. First,
within the SM, the CP asymmetry in $\bd(t) \to J/\psi \ks$ should be
approximately equal to that in decays dominated by the quark-level
penguin transition $\btos q{\bar q}$ ($q=u,d,s$). However, there is a
$2.2\sigma$ difference between the Belle measurement of the CP
asymmetry in $\bd(t) \to \phi\ks$ and that in $\bd(t) \to J/\psi \ks$
\cite{Sakai}. In addition, BaBar sees a $3.0\sigma$ discrepancy
between the CP asymmetries measured in $\bd(t)\to \eta'\ks$ and
$\bd(t) \to J/\psi \ks$ \cite{Giorgi}. Second, the latest data on
$B\to\pi K$ decays (branching ratios and various CP asymmetries)
appear to be inconsistent with the SM \cite{BKpidecays}. Third, within
the SM, one expects no triple-product asymmetries in $B \to \phi K^*$
\cite{BVVTP}, but BaBar has measured such an effect at $1.7\sigma$
\cite{BaBarTP}.

It must be emphasized that none of these signals is statistically
significant. Furthermore, the two experiments Belle and BaBar have not
yet converged on any of the above measurements. Still, these signals
are intriguing, particularly since the decays do share one thing in
common: they all receive significant contributions from $\btos$
penguin amplitudes. If NP is indeed present, it is therefore plausible
to suspect that it is the $\btos$ penguin which is principally
affected. This is the assumption made in this paper.

All new-physics effects in $B$ decays are necessarily virtual. Thus,
regardless of how evidence for physics beyond the SM is found --- any
of the above hints could give a statistically-significant signal of NP
with more data --- many models can explain any discrepancy. As a
result, it has generally been assumed that the {\it identification} of
the NP will have to wait until the new particles are produced directly
at high-energy colliders.

However, it was recently shown that this is not entirely true
\cite{DLNP}. Briefly, the argument is as follows. Following the
experimental hints, we assume that new physics contributes
significantly to those decays which have large $\btos$ penguin
amplitudes. Consider now a $B\to f$ decay involving a $\btos$
penguin. The NP operators are assumed to be roughly the same size as
the SM $\btos$ penguin operators, so the new effects are sizeable. At
the quark level, the NP contributions take the form ${\cal O}_{\sss
NP}^{ij,q} \sim {\bar s} \Gamma_i b \, {\bar q} \Gamma_j q$ ($q =
u,d,s,c$), where the $\Gamma_{i,j}$ represent Lorentz structures, and
colour indices are suppressed. There are a total of 20 possible NP
operators; each of them can in principle have a different weak phase.

There are new-physics contributions to the decay $B\to f$ through the
matrix elements $\bra{f} {\cal O}_{\sss NP}^{ij,q} \ket{B}$. Each of
these can be written as
\beq
\bra{f} {\cal O}_{\sss NP}^{ij,q} \ket{B} = A_k e^{i \phi_k^q} e^{i
\delta_k^q} ~,
\eeq
where $\phi_k^q$ and $\delta_k^q$ are the NP weak and strong phases
associated with the individual matrix elements. However, it was argued
in Ref.~\cite{DLNP} that all NP strong phases are negligible compared
to those of the SM. The point is that strong phases arise from
rescattering. In the SM, this comes mainly from the tree diagram
described at the quark level by $\btos c {\bar c}$~\footnote{There is
also a $\btos u {\bar u}$ tree diagram, but it is described by the
product of CKM matrix elements $V_{ub}^* V_{us}$, which is tiny.}.
However, note that this diagram is quite a bit larger than the $\btos$
penguin diagrams. That is, the strong phases associated with penguin
amplitudes are due to rescattering from a diagram which is
considerably bigger. However, the NP strong phases can come only from
rescattering from the NP diagrams themselves, which are much smaller
than the SM tree diagram. Thus, the generated NP strong phases are
correspondingly smaller than their SM counterparts. That is, the NP
strong phases are negligible compared to the SM strong phases. (A
detailed discussion of small NP strong phases is presented in the
Appendix.)

Note that, in certain calculations of nonleptonic decays \cite{PQCD}
it is claimed that the rescattering from the tree diagrams is
negligible, but that annihilation terms, which are subleading ($\sim
O(1/m_b)$), can be large. Large rescattering from the annihilation
terms can generate a significant strong phase. If this scenario is
true, then annihilation topologies associated with new-physics
operators can also generate a large strong phase through rescattering.
However, there is no general agreement on the importance of
annihilation terms. Ultimately the size of annihilation diagrams is an
experimental question, and can be tested by the measurement of decays
such as $\bd \to D_s^+ D_s^-$ and $\bd\to K^+ K^-$. If the
annihilations terms turn out to be small, as expected from the
$O(1/m_b)$ suppression, then we can neglect the rescattering phase
resulting from them. In our analysis we assume that annihilation-type
topologies in the SM and with NP, which are power suppressed, are
small, and therefore our argument that NP strong phases are negligible
compared to those of the SM remains valid.

The observation that the NP strong phases are negligible allows for a
great simplification: one can now combine all NP matrix elements into
a single NP amplitude, with a single weak phase:
\beq
\sum \bra{f} {\cal O}_{\sss NP}^{ij,q} \ket{B} = \ANPq e^{i \Phi_q} ~,
\eeq
where $q=u,d,s,c$. Throughout the paper, we use the symbols ${\cal A}$
and $\Phi$ to denote the NP amplitudes and weak phases, respectively.
In the above, 
\bea
\tan{ \Phi_q} & = & \frac{ \sum_i A_i \sin{\phi_i^q}} {\sum_i A_i
\cos{\phi_i^q}} ~.
\label{phase_qq}
\eea
Thus, all NP effects can be parametrized in terms of a small number of
NP quantities. That is, we have an effective-lagrangian approach to
new physics in CP violation in the $B$ system.

The argument the new-physics strong phases are negligible is quite
general and applies to all NP models. However, this result can be
obviated if special conditions are met. In particular, it does not
hold if the NP is quite light, or if there is a significant
enhancement of certain matrix elements. While these are disfavoured
theoretically, the reader should be aware of these possible
exceptions.

Note that, in general, $\ANPq$ and $\Phi_q$ will be process-dependent.
The NP phase $\Phi_q$ will be the same for all decays governed by the
quark-level process $\btos q {\bar q}$ only if all NP operators for
the same quark-level process have the same weak phase. This is not
uncommon. There are a number of NP models for which the weak phase is
{\it universal} to all operators. These include models with $Z$-
\cite{ZFCNC} and $Z'$-mediated \cite{Z'FCNC} flavour-changing neutral
currents (FCNC's), models in which the gluonic penguin operators have
an enhanced chromomagnetic moment \cite{chromo}, and models with
scalar-mediated FCNC's \cite{LeeGeorgi}. On the other hand, there are
also NP models without universal weak phases, such as supersymmetric
models with $R$-parity breaking, left-right symmetric models and
models with four generations.

In Ref.~\cite{DLNP}, it was shown that the $\ANPq$ and $\Phi_q$ can be
{\it measured} using pairs of $B$ decays which are related by flavour
SU(3). One decay has a large $\btos$ penguin component and so receives
new-physics contributions. The other has a $\btod$ penguin
contribution. At present, there are no NP signals in processes which
receive sizeable contributions from $\btod$ penguin amplitudes, such
as $\bd\to \pi\pi$. In this technique, and in others like it, we
therefore assume that the NP does not affect decays involving $\btod$
penguins. The measurements of the two decays permit the extraction of
the NP parameters, which in turn allows one to discriminate among NP
models and rule out many of them. We can thus partially identify the
new physics, before high-energy colliders are used.

In this paper, we provide a more detailed description of the method
proposed in Ref.~\cite{DLNP} to measure the NP parameters. We also
examine two additional methods which can be used to obtain these NP
parameters.

The first new method involves $B\to \pi K$ and $B\to\pi\pi$ decays.
Recently, it was shown that, within the SM, the full unitarity
triangle can be extracted from measurements of $B\to \pi K$ decays
\cite{IPLL}. In order to do this, it is necessary to use flavour SU(3)
to relate electroweak penguin operators to tree operators. On the
other hand, if one assumes in addition the presence of new-physics
$\btos$ amplitudes in $B\to \pi K$, it is straightforward to show that
there is not sufficient information to extract the various SM and NP
parameters. However, flavour SU(3) also relates $B\to \pi K$ to
$B\to\pi\pi$ decays. Since the NP is not expected to affect these
latter decays, one can use flavour SU(3) to obtain certain SM $B\to
\pi K$ amplitudes from $B\to\pi\pi$. With this information, it is
possible to {\it measure} the NP parameters. The advantage of this
method is that the analysis can be performed with present data; the
disadvantage is that there is a theoretical error due to the
assumption of flavour SU(3) symmetry.

The second new method involves $\bvv$ decays, where $V_1$ and $V_2$
are vector mesons. These decays are very promising for finding
evidence of physics beyond the SM. Suppose that the final state is
such that (i) ${\overline{V}}_1 {\overline{V}}_2 = V_1 V_2$, and (ii)
a single decay amplitude dominates in the SM. In this case, a
time-dependent angular analysis of $B(t) \to V_1 V_2$ provides
numerous signals of new physics \cite{LSSbounds}. Suppose further that
a single NP amplitude is present, with a different weak phase from
that of the SM amplitude and a (helicity-dependent) strong phase. The
NP weak phase is assumed to be helicity-independent, which is the case
for NP models whose weak phase is universal to all operators. In
Ref.~\cite{LSSbounds} it was shown that one can place lower bounds on
the NP parameters. However, we have argued above that the NP strong
phase is negligible, in which case the analysis is modified. As we
will see, there are now more observables than theoretical parameters,
so that one can {\it measure} the NP parameters in this
system. Compared to Ref.~\cite{DLNP}, the advantage is that no
theoretical input [flavour SU(3)] is required; the disadvantage is
that it is difficult experimentally, and the analysis only holds for a
certain class of NP models.

We begin in Sec.~2 with a detailed discussion of the method involving
$\btos$ and $\btod$ penguin decays which are related by SU(3). In
Sec.~3 we turn to the analysis of $B\to \pi K$ and $B\to\pi\pi$
decays. The technique involving $\bvv$ decays is examined in Sec.~4.
In Sec.~5 we discuss all three methods, stressing their advantages and
disadvantages. All methods have their own unique features, and a
complete analysis would ideally include all three techniques. We also
examine two models of new physics, and show that different NP models
lead to different patterns of NP parameters. This demonstrates that
the measurement of the NP parameters does indeed discriminate among
various models, and provides a partial identification of the NP. We
conclude in Sec.~6.

\section{\boldmath $B$ Penguin Decays}

We begin with a description of the method proposed in Ref.~\cite{DLNP}
for measuring new-physics parameters. This technique closely resembles
that of Ref.~\cite{Bpenguin}, which two of us (AD, DL) recently
proposed for extracting CP phase information. Here the method is
reversed. We assume that NP is present only in decays with large
$\btos$ penguin amplitudes, and that the SM CP phase information is
known (the SM phases can be measured using processes which do not
involve large $\btos$ penguin amplitudes). We first study the general
case in abstract terms. We then apply the method to specific decays.

\subsection{General Case}

We begin by considering a neutral $B^{\prime 0} \to M'_1 M'_2$ decay
in which the final state $M'_1 M'_2$ is accessible to both $B^0$ and
$\Bbar$ mesons. $B^{\prime 0}$ can be either a $\bd$ or a $\bs$ meson,
and $M'_1$ and $M'_2$ are two mesons. (If both $M'_1$ and $M'_2$ are
vector mesons, the final state can be considered as a single helicity
state of $M'_1 M'_2$.) This decay involves a $\btos$ penguin
contribution and is dominated by a single decay amplitude in the SM.
(The case in which there are two significant SM amplitudes is
discussed at the end of this subsection.)

Since $B^{\prime 0} \to M'_1 M'_2$ involves a $\btos$ penguin
amplitude, new physics is present. As discussed in the introduction,
because the NP strong phases are negligible, the effect of the NP can
be parametrized in terms of a single effective amplitude, with a NP
weak phase. Thus, including the NP, the amplitude for $B^{\prime 0}
\to M'_1 M'_2$ can be written
\beq
A(B^{\prime 0} \to M'_1 M'_2) \equiv A = \Actp\ e^{i \delta'_{ct}} +
\ANPq e^{i \Phi_q} ~,
\label{generalNPampdom1}
\eeq
where $\Actp$ and $\ANPq$ are the SM and NP amplitudes,
respectively. Similarly, $\delta'_{ct}$ and $\Phi_q$ are the SM strong
phase and NP weak phase, respectively. The amplitude for the
CP-conjugate process, ${\overline A}$, can be obtained from the above
by changing the sign of the weak phase $\Phi_q$.

The time-dependent measurement of $B^{\prime 0}(t)\to M'_1 M'_2$
allows one to obtain the three observables
\bea
B &\equiv & \frac{1}{2} \left( |A|^2 + |{\overline A}|^2 \right) =
(\Actp)^2 + (\ANPq)^2 + 2 \Actp \, \ANPq \cos\delta'_{ct}
\cos\Phi_q ~, \nn \\
\label{meas1}
a_{dir} &\equiv & \frac{1}{2} \left( |A|^2 - |{\overline A}|^2 \right)
= 2 \Actp \, \ANPq \sin\delta'_{ct} \sin\Phi_q ~, \\
\aI &\equiv & {\rm Im}\left( e^{-2i \phi^q_{\sss M}} A^* {\overline A}
\right) = -(\Actp)^2 \sin 2\phi^q_{\sss M} - 2 \Actp \, \ANPq
\cos\delta'_{ct} \sin (2 \phi^q_{\sss M} + \Phi_q) \nn\\
& & \hskip2.6truein
- (\ANPq)^2 \sin (2\phi^q_{\sss M} + 2 \Phi_q)~. \nn
\eea
It is useful to define a fourth observable:
\bea
\aR & \equiv & {\rm Re}\left( e^{-2i \phi^q_{\sss M}} A^* {\overline A}
\right) = (\Actp)^2 \cos 2\phi^q_{\sss M} + 2 \Actp \, \ANPq
\cos\delta'_{ct} \cos (2 \phi^q_{\sss M} + \Phi_q) \nn\\
& & \hskip2.6truein
+ (\ANPq)^2 \cos (2\phi^q_{\sss M} + 2 \Phi_q)~.
\label{meas2}
\eea
The quantity $\aR$ is not independent of the other three observables:
\beq
{\aR}^2 = {B}^2 - {a_{dir}}^2 - {\aI}^2 ~.
\label{aRdefNP}
\eeq
Thus, one can obtain $\aR$ from measurements of $B$, $a_{dir}$ and
$\aI$, up to a sign ambiguity.

In the above, $\phi^q_{\sss M}$ ($q=d,s$) is the phase of $B^{\prime
0}$--${\bar B}^{\prime 0}$ mixing. For $B^{\prime 0} = \bd$, this
phase is unaffected by new physics and thus takes its SM value,
$\beta$. The canonical way to measure this angle is via CP violation
in $\bd(t)\to J/\psi \ks$. However, there is a potential problem here:
this decay receives NP contributions from $O_{\sss NP}^c \sim \bar{s}b
{\bar c} c$ operators (the Lorentz and colour structures have been
suppressed). On the other hand, the value of $\beta$ extracted from
$\bd(t)\to J/\psi \ks$ is in line with SM expectations. This strongly
suggests that any $O_{\sss NP}^c$ contributions to this decay are
quite small.

The situation is somewhat different for $B^{\prime 0} = \bs$. In
general, NP which affects $\btos$ transitions will also contribute to
$\bs$--$\bsbar$ mixing, i.e.\ one will have NP operators of the form
$\bar{s} b \bar{s} b$. In this case, the phase of $\bs$--$\bsbar$
mixing may well differ from its SM value ($\simeq 0$) due to the
presence of NP. The standard way to measure this mixing phase is
through CP violation in $\bs(t)\to J/\psi \eta$ (or $\bs(t)\to J/\psi
\phi$). As with $\bd(t)\to J/\psi \ks$, these decays potentially
receive $O_{\sss NP}^c$ contributions. However, since the non-strange
part of the $\eta$ wavefunction has a negligible contribution to
$\bra{J/\psi \eta}O_{\sss NP}^c \ket{\bs}$, this matrix element can be
related by flavour SU(3) to $\bra{J/\psi \ks}O_{\sss NP}^c \ket{\bd}$
(up to a mixing angle). That is, both matrix elements are very
small. In other words, we do not expect significant $O_{\sss NP}^c$
contributions to $\bs(t)\to J/\psi \eta$, and the phase of
$\bs$--$\bsbar$ mixing can be measured through CP violation in this
decay, even in the presence of NP.

Another decay which can be used to measure the phase of
$\bs$--$\bsbar$ mixing is $\bs(t) \to D_s^+ D_s^-$. Since the final
state $D_s^+ D_s^-$ is unrelated to $J/\psi \eta$, it is logically
possible that $O_{\sss NP}^c$ will have measurable effects in $\bs(t)
\to D_s^+ D_s^-$. The easiest way to detect this is to measure the
$\bs$--$\bsbar$ mixing phase in both $\bs(t)\to J/\psi \eta$ and
$\bs(t) \to D_s^+ D_s^-$. If these two phases differ, this will
clearly signal the presence of NP in $\btos c {\bar c}$
transitions. However, such measurements will not allow us to cleanly
determine the magnitude and phase of $O_{\sss NP}^c$. This can be done
using the technique described in this section.

Note that the expressions for $a_{dir}$ and $\aI$ in Eq.~(\ref{meas1})
provide several clear signals of NP. Since $B^{\prime 0} \to M'_1
M'_2$ is dominated by a single decay in the SM, the direct CP
asymmetry is predicted to vanish. Furthermore, in the SM the indirect
CP asymmetry measures the mixing phase $\phi^q_{\sss M}$. Thus, if it is
found that $a_{dir} \ne 0$, or that $\phi^q_{\sss M}$ differs from its
SM value, this would be a smoking-gun signal of NP. Note also that, if
it happens that the SM strong phases are small, $a_{dir}$ may be
unmeasurable. In this case, a better signal of new physics is the
measurement of T-violating triple-product correlations in the
corresponding vector-vector final states \cite{BVVTP}. This is an
example of the many NP signals present in $B$ decays. However, these
signals do not, by themselves, allow the measurement of the NP
parameters.

The three independent observables of Eqs.~(\ref{meas1}) and
(\ref{meas2}) depend on four unknown theoretical parameters: $\ANPq$,
$\Actp$, $\delta'_{ct}$ and $\Phi_q$. Therefore one cannot obtain
information about the new-physics parameters $\ANPq$ and $\Phi_q$
from these measurements. However, one can partially solve the
equations to obtain
\beq
(\Actp)^2 = { \aR \cos(2\phi^q_{\sss M} + 2\Phi_q) - \aI
\sin(2\phi^q_{\sss M} + 2\Phi_q) - B \over \cos 2\Phi_q - 1} ~.
\label{phicond}
\eeq
{}From this expression, we see that, if we knew $\Actp$, we could
solve for $\Phi_q$.

In order to get $\Actp$ we consider the partner process $B^0 \to M_1
M_2$ involving a $\btod$ penguin amplitude. In the SM this decay is
related by SU(3) symmetry to $B^{\prime 0} \to M'_1 M'_2$. (In some
cases, this relation only holds if one neglects annihilation- or
exchange-type diagrams \cite{Bpenguin}, which are expected to be
small.) $B^0$ can be either a $\bd$ or a $\bs$ meson and, as with
$B^{\prime 0} \to M'_1 M'_2$, it is assumed that both $B^0$ and ${\bar
B}^0$ can decay to the final state $M_1 M_2$. The partner process can
be a pure penguin decay, or can involve both tree and (non-negligible)
penguin contributions.

Since $\btos$ transitions are not involved, the amplitude for $B^0 \to
M_1 M_2$ receives only SM contributions, and is given by
\bea
A(B^0 \to M_1 M_2) & = & A_u V_{ub}^* V_{ud} + A_c V_{cb}^*
V_{cd} + A_t V_{tb}^* V_{td} \nn\\
& = & (A_u - A_t) V_{ub}^* V_{ud} + (A_c - A_t) V_{cb}^* V_{cd} \nn\\
& \equiv & \Aut\ e^{i \gamma} e^{i \delta_{ut}} + \Act\ e^{i
  \delta_{ct}} ~,
\label{Bfamp}
\eea
where $\Aut \equiv |(A_u - A_t) V_{ub}^* V_{ud}|$, $\Act \equiv |(A_c
- A_t) V_{cb}^* V_{cd}|$, and we have explicitly written the strong
phases $\delta_{ut}$ and $\delta_{ct}$, as well as the weak phase
$\gamma$. In the above, we adopt the $c$-quark convention
\cite{cquarkconv}, in which CKM unitarity is used to eliminate the
$t$-quark term.

As with $B^{\prime 0} \to M'_1 M'_2$, the time-dependent measurement
of $B^0(t)\to M_1 M_2$ allows one to obtain three independent
observables [Eqs.~(\ref{meas1}) and (\ref{meas2})]. These observables
depend on five theoretical quantities: $\Act$, $\Aut$, ${\delta}\equiv
{\delta}_{ut} - {\delta}_{ct}$, $\gamma$ and the mixing phase
$\phi^q_{\sss M}$. However, as discussed above, $\phi^q_{\sss M}$ can
be measured independently using processes which are unaffected by new
physics in $\btos$ penguin amplitudes. The weak phase $\gamma$ can be
measured similarly. For example, it can be obtained from $B^\pm \to
DK$ decays \cite{BDK}. Alternatively, the angle $\alpha$ can be
extracted from $B \to \pi\pi$ \cite{Bpipi}, $B\to\rho\pi$
\cite{Brhopi} or $B\to\rho\rho$ decays \cite{Brhorho}, and $\gamma$
can be obtained using $\gamma = \pi - \beta - \gamma$. Given that
these CP phases can be measured independently, the three observables
of $B^0(t)\to M_1 M_2$ now depend on three unknown theoretical
parameters, so that the system of equations can be solved.

In particular, one can obtain $\Act$:
\beq
\Act^2 = { \aR \cos(2\phi^q_{\sss M} + 2\gamma) - \aI \sin(2\phi_{\sss
M} + 2\gamma) - B \over \cos 2\gamma - 1} ~,
\label{gammacondrevisited}
\eeq
where $\aR$, $\aI$ and $B$ are the observables found in $B^0(t)\to M_1
M_2$.

The key point is that, in the SU(3) limit, one has
\beq
\Act=\lambda \Actp ~,
\label{Actrel}
\eeq
where $\lambda = 0.22$ is the Cabibbo angle. With this relation, the
extraction of $\Act$ from $B^0(t)\to M_1 M_2$ yields $\Actp$. Thus,
Eq.~(\ref{phicond}) can be used to solve for the new physics phase
$\Phi_q$. The NP amplitude $\ANPq$ can also be obtained. There is a
theoretical error in the above relation due to SU(3)-breaking
effects. However, various methods were discussed in
Ref.~\cite{Bpenguin} to reduce this SU(3) breaking. All of these
methods are applicable here. In the end, depending on which pair of
processes is used, the theoretical error can be reduced to the level
of 5--15\%.

At this point we can make an important general observation. As noted
in the introduction [Eq.~(\ref{phase_qq})], the new-physics weak phase
$\Phi_q$ depends on the matrix elements of the various NP operators
for the particular process considered ($A_i$), as well as the
corresponding weak phases $\phi_i^q$. Thus, in general, the value of
$\Phi_q$ extracted from two distinct decay pairs with the same
underlying $\btos q \bar{q}$ transition will be different. There are
two reasons for this. First, certain operators which contribute to one
process may not contribute in the same form to another. (For example,
one decay might be colour-suppressed, while the other is
colour-allowed.) Second, even if the same operators are involved (with
the same form) in two $\btos q \bar{q}$ decays, the matrix elements of
the various operators will depend on the final states considered.
Thus,the $A_i$ in Eq.~(\ref{phase_qq}) are process-dependent in
general, and the value of the phase $\Phi_q$ depends on the particular
decay pair used. However, if all NP operators for the quark-level
process $\btos q \bar{q}$ have the same weak phase $\phi^q$, then the
NP phase $\Phi_q$ will be the same for all decays governed by the same
quark-level process. Hence it is important to measure the phase
$\Phi_q$ in more than one pair of processes with the same underlying
quark transition. If the effective phases are different then it would
be a clear signal of more than one NP amplitude, with different weak
phases, in $\btos q \bar{q}$. Furthermore, in some NP models, the
phases for the different underlying quark transitions $\btos q
\bar{q}$ are related, so that the NP phase is independent of the quark
flavour.

In the above method, we have assumed that the decay $B^{\prime 0} \to
M'_1 M'_2$ is dominated by a single decay amplitude in the SM. This is
the case only for the quark-level decays $\btos q{\bar q}$
($q=d,s,c$). However, it is straightforward to adapt this technique to
$\btos u{\bar u}$, for which $B^{\prime 0} \to M'_1 M'_2$ receives
both tree and $\btos$ penguin contributions in the SM. The process
$\bs \to K^+ K^-$ is an example of such a decay. Including the
new-physics contribution, the amplitude for such decays can be written
\beq
A(B^{\prime 0} \to M'_1 M'_2) = \Autp\ e^{i \gamma} e^{i \delta'_{ut}}
+ \Actp\ e^{i \delta'_{ct}} + \ANPu e^{i \Phi_u} ~.
\eeq
Here, assuming that $\gamma$ and the mixing phase $\phi^q_{\sss M}$ are
known, the three independent observables in this decay depend on six
unknown parameters: $\Actp$, $\Autp$, $\delta'_{ut}$, $\delta'_{ct}$,
$\ANPu$ and $\Phi_u$. In this case, in order to solve for the NP
parameters, one needs three pieces of information. These can be
obtained as follows. Measurements of the partner process allow one to
extract $\Act$, $\Aut$ and ${\delta}\equiv {\delta}_{ut} -
{\delta}_{ct}$. We now assume that
\beq
\Act = \lambda \Actp ~,~~ \lambda \Aut = \Autp ~,~~ \delta' = \delta ~,
\label{3assump}
\eeq
where $\delta' \equiv \delta'_{ut} - \delta'_{ct}$. These assumptions
then permit the extraction of $\ANPu$ and $\Phi_u$ from measurements
of $B^{\prime 0} \to M'_1 M'_2$.

However, the theoretical uncertainty here due to SU(3) breaking is
considerably larger than in the case where $B^{\prime 0} \to M'_1
M'_2$ is dominated by a single amplitude in the SM. Not only do we
relate two amplitudes instead of one [Eq.~(\ref{Actrel})], but we also
assume that two strong phases are equal. Thus, the NP parameters
$\ANPu$ and $\Phi_u$ can be obtained in this way, but we expect a
larger theoretical error.

\subsection{Specific Decays}

In Ref.~\cite{Bpenguin}, we showed that there are twelve decay pairs
$B^0 \to M_1 M_2$ and $B' \to M'_1 M'_2$ which can be used to obtain
CP phase information in the SM, with a small theoretical error. (In
fact, there are more, since many of the particles in the final states
can be observed as either pseudoscalar (P) or vector (V) mesons.)
Many of these decay pairs can also be used to measure the new-physics
parameters, assuming that the NP contributes significantly only to the
$\btos$ decays, and that the SM CP phases have already been measured
using non-$\btos$ processes.

As noted earlier, assuming that new-physics strong rescattering is
negligible relative to that of the SM, all NP effects can be
parametrized in terms of the effective NP amplitudes $\ANPq$ and weak
phases $\Phi_q$ ($q=u,d,s,c$), independent of the type of underlying
NP \cite{DLNP}. Of the above twelve decay pairs, seven involve only
neutral $B$-mesons and have final states $M'_1 M'_2$ accessible to
both $B^0$ and $\Bbar$. These can be used to measure the NP parameters
$\ANPq$ and $\Phi_q$ ($q=d,s,c$) using the method outlined in
Sec.~2.1. For $\ANPu$ and $\Phi_u$, we have to use a decay pair in
which $B' \to M'_1 M'_2$ receives both tree and penguin contributions
in the SM. There is one such possibility. Thus, all four sets of NP
parameters can be obtained from measurements of pairs of $B$ decays.

The decay pairs are listed in Table~\ref{summarytable}, along with the
new-physics parameters probed. We have several comments about these.

\begin{table}
\hfil
\vbox{\offinterlineskip
\halign{&\vrule#&
 \strut\quad#\hfil\quad\cr
\noalign{\hrule}
height2pt&\omit&&\omit&&\omit&\cr
& NP Parameters && $B^{\prime 0}(t) \to M'_1 M'_2$ && $B^0(t) \to M_1
M_2$ & \cr
height2pt&\omit&&\omit&&\omit&\cr
\noalign{\hrule}
height2pt&\omit&&\omit&&\omit&\cr
& $\Phi_c$, $\ANPc$ && $\bs(t) \to D_s^+ D_s^-$ && $\bd(t) \to D^+
D^-$ & \cr
height2pt&\omit&&\omit&&\omit&\cr
\noalign{\hrule}
height2pt&\omit&&\omit&&\omit&\cr
& $\Phi_s$, $\ANPs$ && $\bd(t) \to\phi K^{*0}$ && $\bs(t) \to \phi
{\bar K}^{*0}$ & \cr
& \omit && $\bs(t) \to \phi \phi$ && $\bs(t) \to \phi {\bar K}^{*0}$ &
\cr
height2pt&\omit&&\omit&&\omit&\cr
\noalign{\hrule}
height2pt&\omit&&\omit&&\omit&\cr
& $\Phi_d$, $\ANPd$ && $\bs(t) \to K^0 {\bar K}^0$ && $\bd(t) \to
\pi^+ \pi^-$ & \cr
& \omit && $\bs(t) \to K^0 {\bar K}^0$ && $\bd(t) \to K^0 {\bar K}^0$
& \cr
& \omit && $\bd(t) \to K^{*0}\rho^0$ && $\bd(t) \to\rho^0\rho^0$ & \cr
& \omit && $\bd(t) \to K^{*0}\rho^0$ && $\bs(t) \to {\bar K}^{*0}
\rho^0$ & \cr
height2pt&\omit&&\omit&&\omit&\cr
\noalign{\hrule}
height2pt&\omit&&\omit&&\omit&\cr
& $\Phi_u$, $\ANPu$ && $\bs(t) \to K^+ K^-$ && $\bd(t) \to \pi^+
\pi^-$ & \cr
height2pt&\omit&&\omit&&\omit&\cr
\noalign{\hrule}}}
\caption{For each set of new-physics parameters $\ANPq$ and $\Phi_q$,
we list the $B$ decays ($B^{\prime 0}(t) \to M'_1 M'_2$) and their
partner processes ($B^0(t) \to M_1 M_2$) which can be used to measure
them.}
\label{summarytable}
\end{table}

There are three reasons why certain decays are written in terms of
vector-vector ($VV$) final states, while others involve
pseudoscalar-pseudoscalar ($PP$) states. First, some decays involve a
final-state $\pi^0$. However, experimentally it will be necessary to
find the decay vertices of the final particles. This is virtually
impossible for a $\pi^0$, and so we always use a $\rho^0$
\cite{pi0refs}. Second, some pairs of decays are related by SU(3) in
the SM only if an $\ssbar$ quark pair is used. Unfortunately, there
are no P's which are pure $\ssbar$. The mesons $\eta$ and $\eta'$ have
an $\ssbar$ component, but they also have significant $(u \bar u)$ and
$(d \bar d)$ pieces. As a result the decays $B' \to M'_1 M'_2$ and
$B^0 \to M_1 M_2$ are not really related by SU(3) in the SM if the
final state involves an $\eta$ or $\eta'$. We therefore consider
instead the vector meson $\phi$ which is essentially a pure $\ssbar$
quark state. Finally, we require that both $B^0$ and ${\bar B}^0$ be
able to decay to the final state. This cannot happen if the final
state contains a single $K^0$ (or ${\bar K}^0$) meson. However, it can
occur if this final-state particle is an excited neutral kaon. In this
case one decay involves $K^{*0}$, while the other has ${\bar K}^{*0}$.
Assuming that the vector meson is detected via its decay to $\ks
\pi^0$ (as in the measurement of $\sin 2\beta$ via $\bd(t) \to J/\psi
K^*$), then both $B^0$ and ${\bar B}^0$ can decay to the same final
state.

Apart from these three restrictions, the final-state particles can be
taken to be either pseudoscalar or vector. Indeed, it will be useful
to measure the NP parameters in modes with $PP$, $PV$ and $VV$
final-state particles, since different NP operators are probed in
these decays. For example, within factorization, certain scalar
operators cannot contribute to $PV$ or $VV$ states if their amplitudes
involve the matrix element $\bra{V} \bar{q} \gamma_{L,R} q
\ket{0}$. In general, the matrix element of a given operator will be
different for the various $PP$, $PV$ and $VV$ final states. Thus, the
measurement of the NP parameters in different modes will provide some
clues as to which NP operators are present.

In addition, if it is found that $\Phi_q$ is different for decays
governed by the same underlying quark-level transition, it will
indicate the presence of more than one NP amplitude, with different
weak phases.

For the NP parameters $\ANPq$ and $\Phi_q$ ($q=d,s,c$), the
theoretical error is due to SU(3) breaking in Eq.~(\ref{Actrel}), and
can be reduced to the range 5--15\% \cite{Bpenguin}. On the other
hand, the only way to measure $\Phi_u$ and $\ANPu$ is to use
$\bs(t) \to K^+ K^-$ and $\bd(t) \to \pi^+ \pi^-$. However, $\bs\to
K^+ K^-$ has both tree and penguin contributions. In order to obtain
$\Phi_u$ and $\ANPu$, it is therefore necessary to make the three
assumptions in Eq.~(\ref{3assump}). In the context of measuring the
angle $\gamma$, the SU(3) breaking in $\bs \to K^+K^-$ and $\bd \to
\pi^+ \pi^-$ was examined in Refs.~\cite{bsKK,beneke}. In the
framework of naive factorization or QCD factorization\cite{BBNS}, it
can be shown that, as long as annihilation-type topologies are small,
the double ratio of amplitudes $\lambda^2 (\Actp/\Autp)/(\Act/\Aut)$
has small SU(3) breaking. However, Eq.~(\ref{3assump}) does not
involve this double ratio of amplitudes -- it involves single
amplitude ratios (and an equality of strong phases). In this case,
even within QCD factorization with small annihilation-type topologies,
there are several sources of SU(3) breaking which are not under total
control. The SU(3) breaking comes from the difference between unknown
$\bs \to K$ and $\bd \to \pi$ form factors, differences in the light
cone distributions of the kaon and the pion, and other subleading but
potentially important unknown soft physics \cite{charming}. As a
result, putting all these SU(3)-breaking effects together, the
theoretical error in the extraction of $\Phi_u$ and $\ANPu$ is
quite a bit larger than for the measurement of the other NP
parameters.

Note that only one pair in Table~\ref{summarytable} involves only
$\bd$ decays. The others will require the time-dependent measurement
of $\bs$ decays. However, this will be difficult experimentally, as
$\bs$--$\bsbar$ mixing is large. For this reason the decay pair
$\bd(t) \to K^{*0}\rho^0$ and $\bd(t) \to\rho^0\rho^0$ may be the most
promising for measuring NP parameters using this method.

\section{\boldmath $B\to \pi K$ and $B\to\pi\pi$ Decays}

In this section we consider $B\to \pi K$ and $B\to\pi\pi$ decays. It
is well known that it is possible to express the amplitudes for $B$
decays to two pseudoscalars in terms of a number of distinct SU(3)
operators. This is equivalent to a description in terms of diagrams
\cite{diagrams}. Neglecting the exchange- and annihilation-type
diagrams, which are expected to be small for dynamical reasons, but
including electroweak penguin contributions (EWP's), there are five
diagrams \cite{su3}: (1) a colour-favored tree amplitude $T$ (or
$T'$), (2) a colour-suppressed tree amplitude $C$ (or $C'$), (3) a
gluonic penguin amplitude $P$ (or $P'$), (4) a colour-favored
electroweak penguin amplitude $\pew$ (or $\pewp$), and (5) a
colour-suppressed electroweak penguin amplitude $\pewc$ (or
$\pewcp$). In the following, we denote all diagrams contributing to
$\btod$ ($\btos$) decays without (with) primes.

As described in Sec.~2, the penguin diagram actually contains several
pieces:
\bea
P & = & P_u \, V_{ub}^* V_{ud} + P_c \, V_{cb}^* V_{cd} + P_t \,
V_{tb}^* V_{td} \nn\\
& = & (P_u - P_t) V_{ub}^* V_{ud} + (P_c - P_t) V_{cb}^* V_{cd} \nn\\
& \equiv & P_{ut} \, e^{i\gamma} + P_{ct} ~.
\eea
In the above, we have explicitly written the weak phase $\gamma$; the
amplitudes implicitly include the strong phases. As in Sec.~2, we have
adopted the $c$-quark convention \cite{cquarkconv} in passing from the
first line to the second. For $\btos$ decays, we can write $P'$
analogously to the above, except that we expect $|P'_{ut}| \ll
|P'_{ct}|$ since $\left\vert {V_{ub}^* V_{us} / V_{cb}^* V_{cs}}
\right\vert \simeq 2\%$. In Sec.~2 we neglected the $P'_{ut}$ term. We
will eventually do something similar here as well, but for the moment
we keep all terms in the $B\to \pi K$ amplitudes.

For $\btod$ decays, the EWP contributions are expected to be
negligible. However, they are important for $\btos$ transitions
\cite{su3}. It was recently shown that, to a good approximation, the
EWP's can be related to tree operators using Fierz transformations and
SU(3) symmetry \cite{EWPs}. Ignoring exchange- and annihilation-type
diagrams once again, the relations are
\bea
\pewp & = & {3\over 4} {c_9 + c_{10} \over c_1 + c_2} R (T' + C') +
{3\over 4} {c_9 - c_{10} \over c_1 - c_2} R (T' - C') ~, \nn\\
\pewcp & = & {3\over 4} {c_9 + c_{10} \over c_1 + c_2} R (T' + C') -
{3\over 4} {c_9 - c_{10} \over c_1 - c_2} R (T' - C') ~,
\label{EWPrels}
\eea
where the $c_i$ are Wilson coefficients. Here, the weak phases have
been factored out, so these relations include only strong phases. In
the above,
\beq
R \equiv \left\vert {V_{tb}^* V_{ts} \over V_{ub}^* V_{us}}
\right\vert = {1 \over \lambda^2 \sqrt{\rho^2 + \eta^2}} ~,
\eeq
where $\rho$ and $\eta$ are the CKM parameters. (Note that the three
CP phases in the unitarity triangle are functions of $\rho$ and
$\eta$.) Thus, all $B\to PP$ amplitudes can be expressed in terms of
$T$, $C$, $P_{ut}$, $P_{ct}$ (or their $\btos$ equivalents) and the
weak phases.

We now turn to the four $B\to \pi K$ decays $B^+ \to \pi^+ K^0$, $B^+
\to \pi^0 K^+$, $\bd \to \pi^- K^+$ and $\bd \to \pi^0 K^0$. There are
9 measurements that can be made of this system: four branching ratios,
four direct asymmetries, and one indirect asymmetry (in $\bd(t) \to
\pi^0 \ks$). However, assuming that the $P'_{ut}$ term is negligible,
in the SM the amplitudes can be expressed in terms of 7 parameters:
three diagram magnitudes, two relative strong phases, and two CKM
parameters. Thus there is enough information in $B\to \pi K$ decays
to reconstruct the full unitarity triangle \cite{IPLL}.

We now consider new physics. $B\to \pi K$ decays are $\btos$
transitions, so that NP can affect these decays. There are two classes
of NP operators, differing in their colour structure: ${\bar s}_\alpha
\Gamma_i b_\alpha \, {\bar q}_\beta \Gamma_j q_\beta$ and ${\bar
s}_\alpha \Gamma_i b_\beta \, {\bar q}_\beta \Gamma_j q_\alpha$. The
first class of NP operators contributes with no colour suppression to
final states containing ${\bar q}q$ mesons. The second type of
operator can also contribute via Fierz transformations, but there is a
suppression factor of $1/N_c$, as well as additional operators
involving colour octet currents. Similarly, for final states with
${\bar s} q$ mesons, the roles of the two classes of operators are
reversed. We denote by $\ApNPqph$ and $\ApNPCqph$ the sum of NP
operators which contribute to final states involving ${\bar q}q$ and
${\bar s}q$ mesons, respectively. Here, $\Phi'_q$ and $\Phi_q^{\prime
{\sss C}}$ are the NP weak phases; the strong phases are zero. We
stress that, despite the ``colour-suppressed'' index $C$, the
operators $\ApNPCqph$ are not necessarily smaller than the $\ApNPqph$.

Including these NP operators, the $B\to \pi K$ amplitudes can be
written
\bea
A(B^+ \to \pi^+ K^0) &=& P'_{ut} \, e^{i\gamma} + P'_{ct} - {1 \over 3}
\pewcp + \ApNPCdph ~, \nn\\
\sqrt{2} A(B^+ \to \pi^0 K^+) &=& - P'_{ut} \, e^{i\gamma} - P'_{ct} - T'
\, e^{i\gamma} - C' \, e^{i\gamma} - \pewp \nn\\
& & \quad - ~ {2 \over 3}
\pewcp - \ApNPuph + \ApNPdph - \ApNPCuph ~, \nn\\
A(\bd \to \pi^- K^+) &=& - P'_{ut} \, e^{i\gamma} - P'_{ct} - T'
\, e^{i\gamma} - {2 \over 3} \pewcp - \ApNPCuph ~, \nn\\
\sqrt{2} A(\bd \to \pi^0 K^0) &=& P'_{ut} \, e^{i\gamma} + P'_{ct} - C'
\, e^{i\gamma} - \pewp - {1 \over 3} \pewcp \nn\\
& & \quad - ~ \ApNPuph + \ApNPdph + \ApNPCdph ~.
\eea
Here, each of $P'_{ut}$, $P'_{ct}$, $T'$ and $C'$ include a
(different) strong phase. Note that $\ApNPuph$ and $\ApNPdph$ always
appear in the same combination above. We therefore define $\ApNPcomb
\equiv - \ApNPuph + \ApNPdph$. It is not possible to distinguish the
two component amplitudes.

In the presence of NP, the amplitudes can be written in terms of 16
theoretical quantities: 7 amplitude magnitudes ($|P'_{ut}|$,
$|P'_{ct}|$, $|T'|$, $|C'|$, $|{\cal A}^{\prime, comb}|$, $|{\cal
A}^{\prime {\sss C}, d}|$ and $|{\cal A}^{\prime {\sss C}, u}|$), 4
relative strong phases, 2 SM weak phases, and 3 NP weak phases. (In
the following, we generically refer to the strong phases, SM weak
phases and NP weak phases as ``$\delta$'', ``$\phi$'' and ``$\Phi$,''
respectively.)  Since we have 16 parameters and only 9 measurements,
it is clear that we cannot measure the NP parameters using $B \to \pi
K$ alone. It does not help to make the approximation that $P'_{ut}$ is
negligible.

Important information can be obtained from measurements of the
$B\to\pi\pi$ system. As per our assumptions, new physics does not
affect such decays. Neglecting EWP contributions, which are expected
to be small, the $B \to \pi \pi$ amplitudes can be written
\bea 
\sqrt{2}A(B^+ \to \pi^+ \pi^0) &=& - T \, e^{i\gamma} - C \, e^{i\gamma} ~,
\nn\\
A(\bd \to \pi^+ \pi^-) &=& - T \, e^{i\gamma} - P_{ut} \, e^{i\gamma} -
P_{ct} ~, \nn\\
\sqrt{2} A(\bd \to \pi^0 \pi^0) &=& - C \, e^{i\gamma} + P_{ut}
\, e^{i\gamma} + P_{ct} ~.
\eea
The indirect asymmetry in $\bd(t) \to \pi^+\pi^-$ also involves the
phase of $\bd$--$\bdbar$ mixing, $\beta$, so the 6 $B\to\pi\pi$
measurements [three branching ratios, two direct asymmetries, and one
indirect asymmetry (in $\bd(t) \to \pi^+ \pi^-$)] are a function of 7
theoretical parameters. However, if one assumes that $\beta$ is
measured in $\bd(t) \to J/\psi \ks$, all theoretical quantities can be
extracted \cite{Charles}.

Within flavour SU(3) symmetry, the $B\to \pi\pi$ amplitudes are
related to those in $B\to \pi K$:
\beq
{T' \over T} = {C' \over C} = {P'_{ut} \over P_{ut}} = {\lambda \over
1 - \lambda^2 /2} ~,~~ {P_{ct} \over P'_{ct}} = {\pew \over \pewp} =
{\pewc \over \pewcp} = - {\lambda \over 1 - \lambda^2 /2} ~,
\label{su3rels}
\eeq
where the various amplitudes include the strong phases. With these
relations, we can combine the information obtained in $B \to \pi \pi$
and $B\to \pi K$ decays. As detailed above, there are a total of 16
theoretical parameters. However, there are now also 16 experimental
measurements: 6 in $B\to\pi\pi$, 9 in $B\to \pi K$ and the extraction
of $\beta$ in $\bd(t) \to J/\psi \ks$. It is therefore in principle
possible to solve for all theoretical unknowns, and we would thus {\it
measure} the NP parameters.

Unfortunately, solving 16 nonlinear equations in 16 unknowns will lead
to a large number of discretely-ambiguous solutions. When one adds the
experimental errors, the solutions will be smeared out, and the values
of the NP parameters will essentially remain unknown. To remedy this,
we adopt the procedure of Ref.~\cite{CGRS} in order to reduce the
number of theoretical unknowns. First, we neglect the $P'_{ut}$ term
in the decays $B^+ \to \pi^+ K^0$ and $B^+ \to \pi^0 K^+$. Second, we
remove the dependence on $P_{ut}$ by redefining $T$ and $C$:
\beq
{\tilde T} = T + P_{ut} ~,~~ {\tilde C} = C - P_{ut} ~,
\eeq
with similar redefinitions for the primed quantities. Finally, the
relations in Eq.~(\ref{EWPrels}) no longer hold when ${\tilde T}'$ and
${\tilde C}'$ are used. We therefore neglect the amplitude $\pewcp$.
With this, there is a single relation between $\pewp$ and the tree
diagrams:
\beq
\pewp = {3\over 2} {c_9 + c_{10} \over c_1 + c_2} R ({\tilde T}' +
{\tilde C}') ~.
\label{newEWPrel}
\eeq
Since $|P'_{ut}|,|\pewcp| \ll |P'_{ct}|$ we expect the error
associated with these approximations to be small. Indeed, in
Ref.~\cite{CGRS}, SM fits both with and without the approximations
were performed, and little difference was found.

With the above approximations, the $B\to \pi K$ amplitudes take the
form
\bea
A(B^+ \to \pi^+ K^0) &=& P'_{ct} + \ApNPCdph ~, \nn\\
\sqrt{2} A(B^+ \to \pi^0 K^+) &=& - P'_{ct} - {\tilde T}' \, e^{i\gamma}
- {\tilde C}' \, e^{i\gamma} - \pewp + \ApNPcomb - \ApNPCuph ~, \nn\\
A(\bd \to \pi^- K^+) &=& - P'_{ct} - {\tilde T}' \, e^{i\gamma} - \ApNPCuph
~, \nn\\
\sqrt{2} A(\bd \to \pi^0 K^0) &=& P'_{ct} - {\tilde C}' \, e^{i\gamma} -
\pewp + \ApNPcomb + \ApNPCdph ~,
\eea
while those for $B\to \pi\pi$ are
\bea 
\sqrt{2}A(B^+ \to \pi^+ \pi^0) &=& - {\tilde T} \, e^{i\gamma} - {\tilde
C} \, e^{i\gamma} ~, \nn\\
A(\bd \to \pi^+ \pi^-) &=& - {\tilde T} \, e^{i\gamma} - P_{ct} ~, \nn\\
\sqrt{2} A(\bd \to \pi^0 \pi^0) &=& - {\tilde C} \, e^{i\gamma} + P_{ct} ~.
\eea
The amplitudes with tildes are related as in Eq.~(\ref{su3rels}):
\beq
{{\tilde T}' \over {\tilde T}} = {{\tilde C}' \over {\tilde C}} =
{\lambda \over 1 - \lambda^2 /2} ~.
\label{newsu3rel}
\eeq
There are now 14 theoretical unknown quantities, but 16 measurements.
Thus, we can solve for the NP parameters with few discrete
ambiguities. (In practice, one will fit for all parameters.)

Of course, we have assumed perfect SU(3) symmetry in this procedure
[Eqs.~(\ref{su3rels}) and (\ref{newsu3rel})]. However, we know that
there may be significant SU(3)-breaking effects. In Ref.~\cite{CGRS},
it was noted that factorization appears to hold for colour-allowed
tree diagrams, so that $|{\tilde T}'/{\tilde T}| \simeq f_K/f_\pi$,
but that the SU(3) breaking in the other relations would be left to
experiment. In our case, NP is present in all $\btos$ decays, and this
will mask any SU(3)-breaking effects. One possibility is to assume
that the ratios of magnitudes of amplitudes are known, but no
assumption is made about the strong phases. That is, we write
\beq
\left\vert {{\tilde T}' \over {\tilde T}} \right\vert = f_T ~,~~
\left\vert {{\tilde C}' \over {\tilde C}} \right\vert = f_C ~,~~
\left\vert {P_{ct} \over P'_{ct}} \right\vert = f_P ~.
\label{ampratios}
\eeq
The quantities $f_T$, $f_C$ and $f_P$ are calculated using some
theoretical model (e.g.\ QCD factorization \cite{BBNS}), but all
strong phases are taken to be additional theoretical unknowns. The
problem here is that this adds two theoretical quantities to the
procedure (two strong phases in $B\to\pi\pi$ decays), so we once again
have 16 measurements and 16 unknowns. As discussed above, this leads
to a large number of discretely-ambiguous solutions. For this reason,
it is probably best to assume that the strong phases of primed and
unprimed amplitudes are equal, as with perfect SU(3) symmetry, and
that the magnitude ratios are given by Eq.~(\ref{ampratios}). In this
case the above procedure will yield the NP parameters, but with
sizeable theoretical errors.

Above, we have concentrated on $B\to PP$ decays, where $P$ is a
pseudoscalar. However, the analysis holds equally for $B\to VV$
decays ($V$ is a vector meson). In this case, an angular analysis must
be performed. Note that we have argued that exchange- and
annihilation-type contributions to the $B\to PP$ decays are expected
to be negligible. However, in some approaches to hadronic $B$ decays,
such amplitudes may be chirally enhanced if there are pseudoscalars in
the final state \cite{PQCD,BBNS}. On the other hand, such chiral
enhancements are not present for $VV$ final states, so this is a
potential point in favour of $B\to VV$ decays. It has been recently
claimed that annihilation terms can also be big in certain $B\to VV$
decays in spite of the lack of chiral enhancement \cite{kagan}. As
noted earlier, ultimately, the size of exchange and annihilation
diagrams is an experimental question, and can be tested by the
measurement of decays such as $\bd \to D_s^+ D_s^-$ and $\bd\to K^+
K^-$.

One can even apply the method to $PV$ final states, but things are
considerably more complicated in this case since the NP contributes
differently to the $PV$ and $VP$ final states. In this case, one must
correspondingly increase the number of NP parameters: there are now a
total of 30 parameters in e.g.\ $B \to \pi K^*$ and $B \to \rho K$
decays. These parameters are to be fitted to 32 measurements (9 in $B
\to \pi K^*$, 9 in $B \to \rho K$, 13 in $B\to\rho\pi$, $\beta$ from
$\bd(t) \to J/\psi \ks$). Thus, while one can solve for the NP
parameters in principle, in practice the analysis is very complicated.

We can do better with the $B\to \pi K$/$B\to\pi\pi$ method if we
perform a semi-model-independent analysis by making an assumption
about the general form of ${\cal A}'$'s. We illustrate this below.

\subsection{Isospin-conserving new physics}

One possibility is to make the general assumption that the new physics
is isospin-conserving. For example, this occurs in NP models in which
the gluonic penguin operators have an enhanced chromomagnetic moment
\cite{chromo}. In this case,
\beq
\ApNPuph = \ApNPdph ~,~~ \ApNPCuph = \ApNPCdph \equiv \ApNPCph ~. 
\label{A'rel}
\eeq
This in turn implies that $\ApNPcomb$ is zero, so that there is just
one ${\cal A}'$ remaining. The $B \to \pi K$ amplitudes now take the
form:
\bea
A(B^+ \to \pi^+ K^0) &=& P'_{ct} + \ApNPCph ~, \nn\\
\sqrt{2} A(B^+ \to \pi^0 K^+) &=& - P'_{ct} - {\tilde T}' \, e^{i\gamma}
- {\tilde C}' \, e^{i\gamma} - \pewp - \ApNPCph ~, \nn\\
A(\bd \to \pi^- K^+) &=& - P'_{ct} - {\tilde T}' \, e^{i\gamma} - \ApNPCph
~, \nn\\
\sqrt{2} A(\bd \to \pi^0 K^0) &=& P'_{ct} - {\tilde C}' \, e^{i\gamma} -
\pewp + \ApNPCph ~.
\eea
There are now only 10 theoretical parameters: 4 amplitude magnitudes,
3 $\delta$'s, 2 $\phi$'s and one $\Phi$. Recall that there are 9
measurements in $B\to\pi K$ alone. We therefore need only one
additional measurement to be able to extract all parameters, including
those related to NP: ${\cal A}^{\prime {\sss C}}$ and $\Phi^{\prime
\sss C}$. Ideally, in order to reduce discrete ambiguities, we would
have two additional measurements. These can come from independent
measurements of the SM phases (the 2 $\phi$'s). Thus, for this
particular type of new physics, we do not need measurements in the
$B\to\pi\pi$ system at all. 

Indeed, it is not necssary to make any assumptions about the absence
of new physics in decays with $\btod$ penguins, though we must assume
that the phase of $\bd$--$\bdbar$ mixing is measured in $\bd(t)\to
J/\psi \ks$. If we assume that NP is also present in $B\to\pi\pi$
decays, these amplitudes take the form
\bea
\sqrt{2}A(B^+ \to \pi^+ \pi^0) &=& - {\tilde T} \, e^{i\gamma} - {\tilde
C} \, e^{i\gamma} + \ANPcomb - \ANPCuph + \ANPCdph ~, \nn\\
A(\bd \to \pi^+ \pi^-) &=& - {\tilde T} \, e^{i\gamma} - P_{ct} - \ANPCuph
~, \nn\\
\label{BpipiNP}
\sqrt{2}A(\bd \to \pi^0 \pi^0) &=& - {\tilde C} \, e^{i\gamma} + P_{ct} +
\ANPcomb + \ANPCdph ~,
\eea
where
\beq
\ANPcomb \equiv - \ANPuph + \ANPdph ~.
\eeq
If the NP is isospin-conserving, this implies that
\beq
\ANPuph = \ANPdph ~,~~ \ANPCuph = \ANPCdph \equiv \ANPCph ~. 
\eeq
In this case, $\ANPcomb$ vanishes, and we are left with only one
${\cal A}$, namely $\ANPCph$.

Assuming perfect SU(3) symmetry, the $B\to \pi K$ and $B\to\pi\pi$
amplitudes are described by a total of 12 theoretical quantities. With
16 measurements, it is possible to extract all parameters, including
those describing the NP in the $\btos$ and $\btod$ transitions.

\section{\boldmath $\bvv$ Decays}

In this section we examine $\bvv$ decays in which ${\overline{V}}_1
{\overline{V}}_2 = V_1 V_2$. We consider decays which are described by
the quark-level transitions ${\bar b} \to {\bar c} c {\bar s}$, ${\bar
b} \to {\bar s} s {\bar s}$, or ${\bar b} \to {\bar s} d {\bar
d}$. Within the SM such decays are dominated by a single weak decay
amplitude, and their weak phase is essentially zero in the standard
parametrization \cite{pdg}. (As noted above, there are two significant
contributions --- the tree and penguin amplitudes --- for decays
described by ${\bar b} \to {\bar s} u {\bar u}$.) Since these are all
$\btos$ transitions, there are new-physics contributions. Note that,
in this method, we make no assumptions about NP in $\btod$ decays.

Suppose that the underlying new-physics model is such that the weak
phase is universal to all NP operators. As discussed in the
introduction, this holds for a large number of NP models. In this
case, the NP weak phase $\Phi_q$ will be helicity-independent. Taking
into account the fact that the NP strong phase is negligible, the
decay amplitude for each of the three possible helicity states may be
written as
\bea
A_\lambda \equiv Amp (\bvv)_\lambda &=& a_\lambda e^{i \delta_\lambda}
+ \ANPqlam e^{i\Phi_q} ~, \nn\\
{\bar A}_\lambda \equiv Amp ({\bar B} \to V_1 V_2)_\lambda &=&
a_\lambda e^{i \delta_\lambda} + \ANPqlam e^{-i\Phi_q} ~,
\label{amps}
\eea
where $a_\lambda$ and $\ANPqlam$ represent the helicity-dependent SM
and NP amplitudes, respectively, the $\delta_\lambda$ are the SM
strong phases, and the helicity index $\lambda$ takes the values
$\left\{ 0,\|,\perp \right\}$. Using CPT invariance, the full decay
amplitudes can be written as
\bea
{\cal A} &=& Amp (\bvv) = A_0 g_0 + A_\| g_\| + i \, A_\perp
g_\perp~, \nn\\
{\bar{\cal A}} &=& Amp ({\bar B} \to V_1 V_2) = {\bar A}_0 g_0 + {\bar
A}_\| g_\| - i \, {\bar A}_\perp g_\perp~,
\label{fullamps}
\eea
where the $g_\lambda$ are the coefficients of the helicity amplitudes
written in the linear polarization basis. The $g_\lambda$ depend only
on the angles describing the kinematics \cite{glambda}. 

Using the above equations, we can write the time-dependent decay rates
as
\beq
\Gamma(\bbarp(t) \to V_1V_2) = e^{-\Gamma t} \sum_{\lambda\leq\sigma}
\Bigl(\Lambda_{\lambda\sigma} \pm \Sigma_{\lambda\sigma}\cos(\Delta M
t) \mp \rho_{\lambda\sigma}\sin(\Delta M t)\Bigr) g_\lambda g_\sigma
~.
\label{decayrates}
\eeq
Thus, by performing a time-dependent angular analysis of the decay
$B(t) \to V_1V_2$, one can measure 18 observables. These are:
\bea
\Lambda_{\lambda\lambda}=\displaystyle
\frac{1}{2}(|A_\lambda|^2+|{\bar A}_\lambda|^2),~~&&
\Sigma_{\lambda\lambda}=\displaystyle
\frac{1}{2}(|A_\lambda|^2-|{\bar A}_\lambda|^2),\nn \\[1.ex]
\Lambda_{\perp i}= -\!{\rm Im}({ A}_\perp { A}_i^* \!-\! {\bar
A}_\perp {{\bar A}_i}^* ),
&&\Lambda_{\| 0}= {\rm Re}(A_\| A_0^*\! +\! {\bar A}_\| {{\bar A}_0}^*
), \nn \\[1.ex]
\Sigma_{\perp i}= -\!{\rm Im}(A_\perp A_i^*\! +\! {\bar A}_\perp
{{\bar A}_i}^* ),
&&\Sigma_{\| 0}= {\rm Re}(A_\| A_0^*\!-\! {\bar A}_\| {{\bar A}_0}^*
),\nn\\[1.ex]
\rho_{\perp i}\!=\! {\rm Re}\!\Bigl(e^{-i\phi^q_{\sss M}} \!\bigl[A_\perp^*
{\bar A}_i\! +\! A_i^* {\bar A}_\perp\bigr]\Bigr),
&&\rho_{\perp \perp}\!=\! {\rm Im}\Bigl(e^{-i\phi^q_{\sss M}}\, A_\perp^*
{\bar A}_\perp\Bigr),\nn\\[1.ex]
\rho_{\| 0}\!=\! -{\rm Im}\!\Bigl(e^{-i\phi^q_{\sss M}}[A_\|^* {\bar A}_0\! +
\!A_0^* {\bar A}_\| ]\Bigr),
&&\rho_{ii}\!=\! -{\rm Im}\!\Bigl(e^{-i\phi^q_{\sss M}} A_i^* {\bar
A}_i\Bigr),
  \label{eq:obs}
\eea
where $i=\{0,\|\}$. As in Sec.~2.1, $\phi^q_{\sss M}$ is the weak
phase factor associated with $B_q^0$--${\bar B}_q^0$ mixing. Note that
the signs of the various $\rho_{\lambda\lambda}$ terms depend on the
CP-parity of the various helicity states. We have chosen the sign of
$\rho_{ii}$ to be $-1$, which corresponds to the final state $\phi
K^*$.

For measuring new-physics parameters, the key point is the
following. There are a total of six amplitudes describing $\bvv$ and
${\bar B} \to V_1 V_2$ decays [Eq.~(\ref{amps})]. At best one can
measure the magnitudes and relative phases of these six
amplitudes. Thus, of the 18 observables, only 11 are
independent. However, these observables are a function of only 11
theoretical parameters\footnote{If the NP weak phase is assumed to
helicity-dependent, then there are more theoretical parameters than
there are measurements, and we cannot solve the system of equations.}:
three $a_\lambda$'s, three $\ANPqlam$'s, $\phi^q_{\sss M}$, $\Phi_q$,
and the three strong phases $\delta_\lambda$. In addition, as
discussed in Sec.~2.1, $\phi^q_{\sss M}$ can be measured
independently, so we effectively have 11 equations in 10 unknowns. The
solution will have discrete ambiguities, but many of these can be
removed using the additional observables.

Thus, as advertised, if new physics is found, it is possible to {\it
measure} the NP parameters via a time-dependent angular analysis of
$\bvv$ decays. To be specific, $\ANPslam$ and $\Phi_s$ can be
extracted from $\bd \to \phi K^{*0}$ or $\bs \to\phi\phi$. The decays
$\bd\to K^{*0} \rho^0$ and $\bs\to K^{*0} {\bar K}^{*0}$ can be used
to measure $\ANPdlam$ and $\Phi_d$. (In the decay $\bd\to K^{*0}
\rho^0$, there is a small theoretical error due to the neglect of the
colour-suppressed $\btos u {\bar u}$ tree contribution.) Finally,
measurements of the decay $\bs \to D_s^{*+} D_s^{*-}$ can be used to
obtain $\ANPclam$ and $\Phi_c$.

Note that this analysis is done within the context of a single $\btos$
$\bvv$ decay. In this case, as in the $B\to \pi K$/$B\to\pi\pi$ method
with isospin-conserving new physics, no assumption is necessary about
the absence of new physics in decays with $\btod$ penguins. The only
assumption needed is that the phase of $B_q^0$--${\bar B}_q^0$ mixing,
which may be affected by NP, can be extracted from $\bd(t)\to J/\psi
\ks$ or $\bs(t)\to J/\psi \eta$.

\subsection{Explicit Solution}

Under the assumption that $\phi^q_{\sss M}$ is known independently, we
can construct an analytic solution (this follows closely the analysis
of Ref.~\cite{LSSbounds}). In terms of the theoretical parameters, the
explicit expressions for the observables are as follows:
\begin{eqnarray}
\Lambda_{\lambda\lambda} & \!\!=\!\! & a_\lambda^2 + (\ANPqlam)^2 + 2
a_\lambda \ANPqlam \cos\delta_\lambda \cos\Phi_q ~, \nn\\
\Sigma_{\lambda\lambda} & \!\!=\!\! & 2 a_\lambda \ANPqlam
\sin\delta_\lambda \sin\Phi_q ~, \nn\\
\Lambda_{\perp i} & \!\!=\!\! & 2 \left[ a_\perp \ANPqi
  \cos\delta_\perp - a_i \ANPqperp \cos\delta_i \right] \sin\Phi_q
~,\nn\\
\Lambda_{\| 0} & \!\!=\!\! & 2 \left[ a_\| a_0 \cos(\delta_\| -
\delta_0) + a_\| \ANPqo \cos\delta_\| \cos\Phi_q + a_0 \ANPqpar
\cos\delta_0 \cos\Phi_q + \ANPqpar \ANPqo \right],\nn\\
\Sigma_{\perp i} & \!\!=\!\! & -2 \left[ a_\perp a_i \sin
(\delta_\perp - \delta_i) + a_\perp \ANPqi \sin\delta_\perp
\cos\Phi_q - ~ a_i \ANPqperp \sin\delta_i \cos\Phi_q \right],
\nn\\
\Sigma_{\| 0} & \!\!=\!\! & 2 \left[ a_\| \ANPqo \sin\delta_\| + a_0
\ANPqpar \sin\delta_0 \right] \sin\Phi_q ~, \nn\\
\rho_{ii} & \!\!=\!\! & a_i^2 \sin 2\phi^q_{\sss M} + 2 a_i \ANPqi
\cos\delta_i \sin(2 \phi^q_{\sss M} + \Phi_q) + (\ANPqi)^2
\sin(2\phi^q_{\sss M} + 2 \Phi_q), \nn\\
\rho_{\perp\perp} & \!\!=\!\! & - a_\perp^2 \sin 2\phi^q_{\sss M} - 2
a_\perp \ANPqperp \cos\delta_\perp \sin(2 \phi^q_{\sss M} + \Phi_q) -
(\ANPqperp)^2 \sin(2\phi^q_{\sss M} + 2 \Phi_q), \nn\\
\rho_{\perp i} & \!\!=\!\! & 2 \left[ a_i a_\perp \cos (\delta_i -
\delta_\perp) \cos 2\phi^q_{\sss M} + a_\perp \ANPqi \cos\delta_\perp
\cos(2 \phi^q_{\sss M} + \Phi_q) \right. \nn\\
& & ~~ \left. + ~ a_i \ANPqperp \cos\delta_i \cos(2 \phi^q_{\sss M}
+ \Phi_q) + ~ \ANPqi \ANPqperp \cos(2\phi^q_{\sss M} + 2\Phi_q)
\right],\nn\\
\rho_{\| 0} & \!\!=\!\! & 2 \left[ a_0 a_\| \cos(\delta_\| - \delta_0)
\sin 2\phi^q_{\sss M} + a_\| \ANPqo \cos\delta_\| \sin(2 \phi^q_{\sss
M} + \Phi_q) \right. \nn\\
& & ~~ \left. + ~ a_0 \ANPqpar \cos\delta_0 \sin(2 \phi^q_{\sss M} +
\Phi_q) + ~ \ANPqo \ANPqpar \sin(2\phi^q_{\sss M} + 2\Phi_q)
\right].
\label{observables}
\end{eqnarray}
For $\bvv$ decays, the analogue of the usual direct CP asymmetry
$a^{CP}_{dir}$ is the helicity-dependent quantity $a_{\lambda}^{dir}
\equiv \Sigma_{\lambda\lambda}/\Lambda_{\lambda\lambda}$. We define
the related quantity
\beq
y_\lambda \equiv \sqrt{1 -
\Sigma_{\lambda\lambda}^2/\Lambda_{\lambda\lambda}^2}~.
\eeq
Similarly, the value of $\sin 2\phi^q_{\sss M}$ measured in $\bvv$
decays can depend on the helicity of the final state:
$\rho_{\lambda\lambda}$ can be recast in terms of a measured weak
phase $(2\phi^q_{\sss M})^{meas}_{\lambda}$, defined as
\bea
\sin\, 2\,(\phi^q_{\sss M})^{meas}_{\lambda} \equiv \frac{\pm
\rho_{\lambda\lambda}}{\sqrt{\Lambda^2_{\lambda\lambda}-
\Sigma^2_{\lambda\lambda}}} ~,
\label{betameasdef}
\eea
where the $+$ $(-)$ sign corresponds to $\lambda=0,\|$ ($ \perp$).

Using the expressions for $\Lambda_{\lambda\lambda}$,
$\Sigma_{\lambda\lambda}$ and $(\phi^q_{\sss M})^{meas}_\lambda$ above,
one can express $a_\lambda$ and $\ANPqlam$ as follows
\cite{LSSbounds}:
\bea
2\,a_\lambda^2\,\sin^2\Phi_q &=& \Lambda_{\lambda\lambda} \left( 
1 - y_\lambda \cos(\theta_\lambda-2\Phi_q) \right) ~, \nn \\
2\,(\ANPqlam)^2\,\sin^2\Phi_q &=& \Lambda_{\lambda\lambda} \left(
1-y_\lambda\cos\theta_\lambda \right) ~,
\eea
where $\theta_\lambda \equiv (2\phi^q_{\sss M})^{meas}_\lambda -
2\phi^q_{\sss M}$. Using these expressions, along with those for
$\Lambda_{\lambda\lambda}$ and $\Sigma_{\lambda\lambda}$
[Eq.~(\ref{observables})], we can solve for $\tan\delta_\lambda$:
\beq
\tan\delta_\lambda=\frac{\Sigma_{\lambda\lambda} \sin 2\Phi_q}
{\Lambda_{\lambda\lambda} \left( -2\cos^2\Phi_q + y_\lambda
\cos(\theta_\lambda - 2\Phi_q) + y_\lambda\cos\theta_\lambda
\right)} ~.
\label{tandelta}
\eeq
This equation expresses $\delta_\lambda$ in terms of observables and
$\Phi_q$.

How one proceeds further depends on which other observables are
available. Suppose that $\Lambda_{\perp i}$ and $\Sigma_{\perp i}$
have been measured. These two observables can be expressed as
\cite{LSSbounds}
\bea
\Sigma_{\perp i} &=& P_i P_\perp \left[ \left( \xi_\perp\,\sigma_i
  -\xi_i\,\sigma_\perp \right) \, \cos\Delta_i -
  \left(1+\xi_i\,\xi_\perp+ \sigma_i\,\sigma_\perp \right) \,
  \sin\Delta_i \right] ~,\nn\\
\Lambda_{\perp i} &=& P_i P_\perp \left[ \left( \xi_\perp-\xi_i
\right) \, \cos\Delta_i - \left( \sigma_i + \sigma_\perp \right)
\sin\Delta_i \, \right]~,
\label{SigLam}
\eea
where $\Delta_i \equiv \delta_\perp-\delta_i$, and
\bea
P_\lambda^2 &\equiv& \Lambda_{\lambda\lambda}(1-y_\lambda\,
       \cos(2\theta_\lambda-2\Phi_q))~, \nn\\
\xi_\lambda &\equiv& \frac{\Lambda_{\lambda\lambda}\, y_\lambda\,
       \sin(2\theta_\lambda-2\Phi_q)}{P_\lambda^2}~, \nn \\
\sigma_\lambda &\equiv& \frac{\Sigma_{\lambda\lambda}}{P_\lambda^2} ~.
\eea
Eqs.~(\ref{SigLam}) can be solved for $\Delta_i$:
\beq
\tan\Delta_i =
\frac{\left(\xi_\perp\,\sigma_i-\xi_i\,\sigma_\perp\right)
\Lambda_{\perp i}-\Sigma_{\perp i}\left(\xi_\perp-\xi_i\right)}
{\left(1+\xi_i\,\xi_\perp+\sigma_i\,\sigma_\perp\right) \Lambda_{\perp
i}-\Sigma_{\perp i}\left(\sigma_i+\sigma_\perp\right)} ~.
\label{tanDeltaone}
\eeq
This expresses $\tan\Delta_i$ in terms of observables and
$\Phi_q$. However, we can also write
\beq
\tan\Delta_i =\frac{\tan\delta_\perp-\tan\delta_i}
{1+\tan\delta_\perp\tan\delta_i} ~.
\label{tanDeltatwo}
\eeq
Eqs.~(\ref{tandelta}), (\ref{tanDeltaone}) and (\ref{tanDeltatwo}) can
then be combined to give a single equation as a function of
$\Phi_q$. This can be solved to get the new-physics weak phase,
which will permit the measurement of the remaining theoretical
parameters.

\subsection{Are the NP strong phases negligible?}

The time-dependent angular analysis also allows us to {\it test} the
assumption that the NP strong phases are negligible. Assume that the
$\bvv$ amplitudes contain (helicity-dependent) NP strong phases
$\Delta_\lambda^q$. In this case Eq.~(\ref{amps}) can be written
\bea
A_\lambda \equiv Amp (\bvv)_\lambda &=& a_\lambda e^{i \delta_\lambda}
+ \ANPqlam e^{i\Phi_q} e^{i \Delta_\lambda^q} ~, \nn\\
{\bar A}_\lambda \equiv Amp ({\bar B} \to V_1 V_2)_\lambda &=&
a_\lambda e^{i \delta_\lambda} + \ANPqlam e^{-i\Phi_q} e^{i
  \Delta_\lambda^q} ~.
\eea
There are now 13 theoretical parameters, and only 11 observables.
However, since the expressions for the observables in terms of
parameters are nonlinear, one can obtain {\it bounds} on the various
theoretical quantities \cite{LSSbounds}. In particular, one can
constrain the two NP strong phase differences $\Delta_\perp^q -
\Delta_i^q$, $i=0,\|$. If either of these bounds is inconsistent with
zero, this will invalidate the assumption of negligible NP strong
phases.

\section{Discussion}

In Ref.~\cite{DLNP} it was argued that the NP strong phases are
negligible relative to those of the SM. In this case, all NP matrix
elements for a given $\btos q {\bar q}$ process ($q=u,d,s,c$) can be
summed into a single effective NP operator, with amplitude $\ANPq$ and
corresponding weak phase $\Phi_q$. In the previous sections, we have
examined three different methods for measuring these NP parameters.
They are: (i) the comparison of $\btos$ and $\btod$ penguin $B$ decays
\cite{DLNP}, (ii) the combined measurement of $B\to \pi K$ and
$B\to\pi\pi$ decays, and (iii) the time-dependent angular analysis of
$\bvv$ decays. All three methods have their particular advantages and
disadvantages.

There are several $B$ decay pairs to which the $B$-penguin method can
be applied. Depending on which pair is used, the $s$-, $d$- and
$c$-quark NP parameters can be obtained with a theoretical error of
about 5--15\%. The $u$-quark NP parameters $\ANPu$ and $\Phi_u$ can
also be measured, but with a much larger theoretical error. A key
assumption is that all SM weak phases have already been measured.
Also, most pairs of modes involve time-dependent $\bs$ decays, which
are hard to measure.

The $B\to \pi K$/$B\to\pi\pi$ method probes only the $d$- and
$u$-quark NP parameters, and with a large ($\gsim 25\%$) theoretical
error. Still, the theoretical error on $\ANPu$ and $\Phi_u$ might well
be smaller than in the method with $B$ penguin decays. The main
advantage of this method is that the analysis can be done now.
Ref.~\cite{CGRS} analyzed these $B$ decays within the SM; it is
straightforward to include NP parameters in the analysis. It is also
possible to adapt the method to include NP effects in $B \to\pi\pi$ if
one makes some assumptions about the nature of the NP.

The $\bvv$ method has no theoretical error -- flavour SU(3) is not
required, and we make no assumption about new physics in $\btod$
transitions. However, this method only allows us to measure the $s$-,
$d$- and $c$-quark NP parameters, and it applies only to those
new-physics models in which the NP weak phase is universal.
Furthermore, it is very difficult experimentally. On the other hand,
it can be used to {\it test} the assumption of negligible NP strong
phases.

At several points previously, we have argued that it is important to
measure all the NP parameters, and in as many different ways a
possible. In the following, we show how different NP models lead to
different patterns of NP parameters. Thus, the measurement of the NP
parameters can rule out certain models and point towards others.

There has already been a great deal of theoretical work discussing
various NP models which can explain the apparent discrepancy in the
Belle measurement of $\sin 2\beta$ in $\bd(t) \to \phi\ks$
\cite{phiKsNPZFCNC,phiKsNPRpar,phiKsNP}. Our aim here is not to
produce an exhaustive analysis of such NP models. Instead we consider
only two, and show that the measurement of the NP parameters can
distinguish between them.

\subsection{\boldmath $Z$-mediated FCNC's}

The first new-physics model we consider is $Z$-mediated (or
$Z'$-mediated) FCNC's \cite{ZFCNC}. This model has received much
attention as a potential explanation of the $\bd(t) \to \phi\ks$
result \cite{phiKsNPZFCNC}. The $Zb{\bar s}$ FCNC coupling which leads
to the $\btos$ transitions is parametrized by the independent
parameter $U_{sb}^{\sss Z}$:
\beq
{\cal L}^{\sss Z}_{\sss FCNC} = - {g \over 2 \cos\theta_{\sss W}} \,
U_{sb}^{\sss Z} \, {\bar s}_\lft \gamma^\mu b_\lft Z_\mu ~.
\eeq
Note that the FCNC involves only left-handed $s$ and $b$ quarks.
These couplings are effectively new contributions to the electroweak
penguin operators of the SM.

The new-physics weak phase arises because $U_{sb}^{\sss Z}$ can be
complex. However, because this parameter is universal, the weak phase
of all NP operators will be the same. This model therefore predicts
the equality of all NP weak phases $\Phi_q$ ($q=u,d,s,c$). If this
condition is not found to be satisfied, the model is ruled out.

\subsection{\boldmath Supersymmetry with $R$-parity breaking}

In supersymmetric (SUSY) models with $R$-parity breaking, the relevant
part of the $R$-parity breaking piece is given by
\beq
{\cal W}_{\not \! R} = \lambda_{ijk}^{\prime}L_iQ_jD_k^c +\frac{1}{2}
\lambda_{i[jk]}^{\prime\prime} U_i^cD_j^cD_k^c ~.
\label{lag}
\eeq
Here $L_i$ and $Q_i$ are the left-handed lepton and quark doublet
superfields, respectively, and $U_i$ and $D_i$ are the left-handed
quark singlet chiral superfields, where $i,j,k$ are generation indices
and $c$ denotes a charge conjugate field.

In the above, the $\lambda^{\prime}$ and $\lambda^{\prime\prime}$
couplings violate lepton number and baryon number, respectively. The
non-observation of proton decay imposes very stringent conditions on
the simultaneous presence of both couplings \cite{Proton}. One
therefore assumes the existence of either $L$-violating couplings or
$B$-violating couplings, but not both.

Now, $\lambda^{\prime\prime}_{i[jk]}$ is antisymmetric in the last two
indices. Thus, there are no $B$-violating couplings which can lead to
the $\btos s {\bar s}$ decay necessary to explain the $\bd(t) \to
\phi\ks$ result of Belle. We will therefore concentrate only on
$L$-violating couplings. (For a discussion of $R$-parity violation and
$\bd(t) \to \phi\ks$, see Ref.~\cite{phiKsNPRpar}.) In terms of
four-component Dirac spinors, the $L$-violating couplings are given by
\cite{DatXin}
\bea
{\cal L}_{\lambda^{\prime}}&=&-\lambda^{\prime}_{ijk}
\left [\tilde \nu^i_L\bar d^k_R d^j_L+\tilde d^j_L\bar d^k_R\nu^i_L
       +(\tilde d^k_R)^*(\bar \nu^i_L)^c d^j_L\right.\nonumber\\
& &\hspace{1.5cm} \left. -\tilde e^i_L\bar d^k_R u^j_L
       -\tilde u^j_L\bar d^k_R e^i_L
       -(\tilde d^k_R)^*(\bar e^i_L)^c u^j_L\right ]+h.c.\
\label{Lviolating}
\eea

{}From this Lagrangian, we see that there are $R$-parity-violating
contributions to all $\btos$ transitions \cite{lambdabNP}. There is a
single contribution to each of the decays $\btos u {\bar{u}}$ and
$\btos c {\bar{c}}$:
\bea
L^u_{eff} & = & -\frac{\lambda^{\prime}_{i12} \lambda^{\prime*}_{i13}}
{2 m_{ \widetilde{e}_i}^2} \bar u_\alpha \gamma_\mu \gamma_L u_\beta
\, \bar s_\beta \gamma_\mu \gamma_R b_\alpha ~, \nn\\
L^c_{eff} & = & -\frac{\lambda^{\prime}_{i22} \lambda^{\prime*}_{i23}}
{2 m_{ \widetilde{e}_i}^2} \bar c_\alpha \gamma_\mu \gamma_L c_\beta
\, \bar s_\beta \gamma_\mu \gamma_R b_\alpha ~.
\eea
For $\btos d{\bar{d}}$, there are four terms:
\bea
L^d_{eff}&=&
\frac{\lambda^{\prime}_{i11} \lambda^{\prime*}_{i23}}{  m_{ \widetilde{\nu}_i}^2}
\bar d \gamma_L d \,
\bar {s}\gamma_R b
+\frac{\lambda^{\prime}_{i32} 
\lambda^{\prime*}_{i11}}{  m_{ \widetilde{\nu}_i}^2}
\bar d \gamma_R d \, \bar s \gamma_L b \nonumber \\
&-&\frac{\lambda^{\prime}_{i12} \lambda^{\prime*}_{i13}}
{2  m_{ \widetilde{\nu}_i}^2}
\bar d_\alpha \gamma_\mu \gamma_L d_\beta \, \bar s_\beta \gamma_\mu 
\gamma_R b_\alpha
-\frac{\lambda^{\prime}_{i31} \lambda^{\prime *}_{i21}}
{2  m_{ \widetilde{\nu}_i}^2}
\bar d_\alpha \gamma_\mu \gamma_R d_\beta \, \bar s_\beta \gamma_\mu \gamma_L 
b_\alpha ~.
\eea
Finally, the relevant Lagrangian for the $\btos s {\bar{s}}$
transition is
\beq
L^s_{eff} = \frac{\lambda^{\prime}_{i32} \lambda^{\prime*}_{i22}} { m_{
\widetilde{\nu}_i}^2} \bar s \gamma_R s \, \bar {s}\gamma_L b+
\frac{\lambda^{\prime}_{i22} \lambda^{\prime*}_{i23}} { m_{
\widetilde{\nu}_i}^2} \bar s \gamma_L s \, \bar {s}\gamma_R b ~.
\eeq

{}From the above expressions we can deduce the following predictions
of $R$-parity-violating SUSY models. First, in general, all four NP
parameters are present and are unrelated to one another. Second, since
there is only a single term contributing to each of $\btos u {\bar u}$
and $\btos c {\bar c}$ transitions, the measured values of $\Phi_u$
and $\Phi_c$ should be independent of the decay pairs considered. On
the other hand, since there is more than one contribution to both
$\btos d {\bar d}$ and $\btos s {\bar s}$, the values of $\Phi_d$ and
$\Phi_s$ will in general be process-dependent. Also, these will differ
from $\Phi_u$ and $\Phi_c$. Should this pattern of NP weak phases not
be found experimentally, we can either rule out or constrain this
model of new physics.

\vskip5truemm

The above two examples illustrate that, indeed, different NP models
lead to different patterns of the NP parameters $\ANPq$ and $\Phi_q$.
Thus, the knowledge of the NP parameters will allow us to discriminate
among various models, and rule certain ones out entirely. In order to
(partially) identify the NP, it will be important to measure its
parameters in as many different ways and decay modes as possible. A
complete analysis of NP parameters would therefore include their
measurement using all three of the methods described in the previous
sections.

\section{Conclusions}

The main purpose of the study of CP violation in the $B$ system is to
look for physics beyond the SM. There are now many theoretical signals
of such new physics, and in fact there are several experimental hints
of NP in decays involving $\btos$ penguin amplitudes. Still, the
conventional thinking was that the {\it identification} of NP could
only be done at future high-energy colliders, where the new particles
could be directly produced. However, recently it was shown that it is
possible to {\it measure} the NP parameters in $B$ decays
\cite{DLNP}. The key observation is that the strong phases associated
with the NP operators are negligible compared to those of the SM. In
this case, all NP matrix elements for a given $\btos q {\bar q}$
process ($q=u,d,s,c$) can be summed into a single effective NP
operator, with amplitude $\ANPq$ and corresponding weak phase
$\Phi_q$. These NP parameters can be measured, allowing a partial
identification of the new physics.

In this paper we have discussed three methods of measuring the NP
parameters. In most cases, it is assumed that the new physics
contributes only to decays with large $\btos$ penguin amplitudes,
while decays involving $\btod$ penguins are not affected. The first
method, initially proposed in Ref.~\cite{DLNP}, employs the comparison
of time-dependent $\btos$ and $\btod$ penguin $B$ decays. The second
uses the combined measurement of $B\to \pi K$ and $B\to\pi\pi$
decays. The third requires the time-dependent angular analysis of
$\bvv$ decays.

The three methods can be used to probe different NP parameters, and
with different theoretical errors. The $B$-penguin method allows us to
obtain the $s$-, $d$- and $c$-quark NP parameters with a theoretical
error of about 5--10\%. The $u$-quark NP parameters can also be
measured, but with a much larger theoretical error. The $B\to \pi
K$/$B\to\pi\pi$ method probes only the $d$- and $u$-quark NP
parameters, but with a large ($\gsim 25\%$) theoretical error. The
$\bvv$ method has no theoretical error, but it applies only to NP
models with a universal weak phase, and it only allows the measurement
of the $s$-, $d$- and $c$-quark NP parameters.

The three methods also have different levels of experimental
difficulty. Most pairs of modes in the $B$-penguin method involve
time-dependent $\bs$ decays, which are hard to measure, and the
time-dependent angular analysis of $\bvv$ decays is very difficult
experimentally. On the other hand, the $B\to \pi K$/$B\to\pi\pi$
method can be performed with present data.

Ideally, a full analysis of NP parameters would use all three methods.
Then all the NP parameters $\ANPq$ and $\Phi_q$ can be measured in
many different ways, and using various decay modes. In general,
different NP models lead to different patterns of the NP operators.
The knowledge of the NP parameters will thus allow us to discriminate
among various models and partially identify the new physics, before
the direct production of new particles at high-energy colliders.

\bigskip
\noindent
{\bf Acknowledgements}:
%\bigskip
The work of A.D., M.I., D.L. and V.P. was financially supported by
NSERC of Canada. 

\newpage
\setcounter{equation}{0}
\def\theequation {A. \arabic {equation}}
\section*{Appendix}

In this Appendix, we present a more detailed discussion of the claim
that the NP strong phases are negligible. To understand our argument,
it is crucial to examine where the SM and NP strong phases come
from. Note that this argument is not dependent on any particular
calculational framework. However, it will be useful to demonstrate it
within a particular approach for calculating nonleptonic decays. We
will follow the method of QCD factorization \cite{BBNS}, also known as
the BBNS approach.

To be more specific let us condider $ B \to K \pi $ decays. The
starting point for calculations of this decay is the SM effective
hamiltonian for $B$ decays \cite{BuraseffH}:
\beq
H_{eff}^q = {G_F \over \protect \sqrt{2}} [V_{fb}V^*_{fq}(c_1O_{1f}^q
+ c_2 O_{2f}^q) - \sum_{i=3}^{10}(V_{tb}V^*_{tq} c_i^t) O_i^q] +
h.c. ~,
\label{Heff}
\eeq
where the superscript $t$ indicates the internal top quark, $f$ can be
the $u$ or $c$ quark, and $q$ can be either a $d$ or $s$ quark. The
operators $O_i^q$ are defined as
\bea
O_{f1}^q &=& \bar q_\alpha \gamma_\mu Lf_\beta\bar
f_\beta\gamma^\mu Lb_\alpha\;,\;\;\;\;\;\;O_{2f}^q =\bar q
\gamma_\mu L f\bar
f\gamma^\mu L b\;,\nn\\
O_{3,5}^q &=&\bar q \gamma_\mu L b
\bar q' \gamma^\mu L(R) q'\;,\;\;\;\;\;\;\;O_{4,6}^q = \bar q_\alpha
\gamma_\mu Lb_\beta
\bar q'_\beta \gamma^\mu L(R) q'_\alpha\;,\\
O_{7,9}^q &=& {3\over 2}\bar q \gamma_\mu L b  e_{q'}\bar q'
\gamma^\mu R(L)q'\;,\;O_{8,10}^q = {3\over 2}\bar q_\alpha
\gamma_\mu L b_\beta
e_{q'}\bar q'_\beta \gamma^\mu R(L) q'_\alpha ~, \nn
\eea
in which $R(L) = 1 \pm \gamma_5$, and $q'$ is summed over $u$, $d$,
$s$, $c$. $O_2$ and $O_1$ are the tree-level and QCD-corrected
operators, respectively. $O_{3-6}$ are the strong gluon-induced
penguin operators, and operators $O_{7-10}$ are due to $\gamma$ and
$Z$ exchange (electroweak penguins), and ``box'' diagrams at loop
level.

First, we consider the SM alone. In particular, we consider the QCD
penguin operators $O_{4,6}$ which contribute dominantly to the SM
penguin amplitude for $B \to K \pi$. (Note that $V_{tb}V^*_{ts}$ in
$H_{eff}$ can be expressed in terms of $V_{cb}^* V_{cs}$ and $V_{ub}^*
V_{us}$ using CKM unitarity.) Consider the $O_{4,6}$ operators with
weak phase $V_{cb}^* V_{cs}$. (The same arguments apply to
contributions with weak phase $V_{ub}^* V_{us}$.) The SM penguin
amplitude can obtain a strong phase from two types of rescattering,
which we call class I and class II:
\begin{itemize}

\item In class I, there is rescattering from the tree-level operators
with Wilson coefficients $C_1 \approx 1$. The operator has the same
weak phase as $O_{4,6}$ and leads to a ${\bar b} \to {\bar c} c {\bar
s}$ transition. This will contribute to a final state like $K \pi$
only through rescattering and will consequently generate a strong
phase. The matrix element in question is $\langle K\pi | O_1 | B
\rangle$, and is called $P_c$. In the BBNS picture this rescattering
is represented by the penguin function $G(x)$ and is proportional to
$C_1 \approx 1$. The $G(x)$ generates the dominant strong phase.

In the usual approach, including that of BBNS, the contribution of the
tree operator is combined with the matrix element of the $O_{4,6}$
operator in the quantity $a_{4,6}$ to represent the SM penguin
contribution. The scattering from the tree operator is the dominant
rescattering effect and is responsible for the strong phase of the SM
penguin amplitude. Note that the strength of the $O_{4,6}$ operators,
$C_{4,6}$, is only $ \sim$ 4\%. Thus, even though rescattering costs
$\alpha_s$, it is enhanced by the large Wilson coefficient $C_1$,
which is 25 times as large as $C_4$. The rescattered tree amplitude is
therefore roughly the same size as the matrix element of the $O_{4,6}$
operator, so that it can generate a significant strong phase in the
BBNS approach \cite{BBNS}.

\item In class II, one can have rescattering from the operators
$O_{4,6}$ themselves: $\langle K\pi | O_{4,6} | B \rangle$. This is
represented by BBNS's $g(x)$ and $G(x)$, and is proportional to $C_4$
or $C_6$. However, as mentioned above, the size of $C_4$ and $C_6$ is
only about 4\%. Unlike Class I rescattering, there are no operators
with large Wilson coefficients to produce an appreciable strong phase
here. In other words, the strength of these operators is much smaller
than $C_1 \approx 1$. So this class of rescattering is subdominant.

\end{itemize}

Now we consider the presence of NP. At $m_b$, one can separate the
effective Hamiltonian into two pieces: $O_4$ (say), which is the SM
operator with the SM Wilson coefficient, and $\widetilde{O}_4$ which
contains all the NP contributions. We assume that $\widetilde{O}_4$
and $O_4$ are of similar size. The strong phase now comes from three
sources: rescattering from $O_1$ (class I), $O_4$ (class II) and
$\widetilde{O}_4$ (class II). The rescattering effects from $O_1$ and
$O_4$ are simply those corresponding to the SM, whose strong phase can
be large. However, the strong phase of the matrix element of
$\widetilde{O}_4$ comes {\it only} from rescattering from the NP
operator itself: $\langle K\pi | \widetilde{O}_4 | B \rangle$. That
is, the NP rescattering is only of the Class II type. Hence, the NP
strong phases are {\it subdominant}. To a first approximation, we
therefore ignore the NP strong phases compared to the SM strong
phases.

To demonstrate this point numerically let us work with a specific
model of new physics, R-parity violating SUSY, that was considered in
Sec. 5. In this model the Lagrangian for $\btos u {\bar{u}}$ and
$\btos c {\bar{c}}$ transitions is :
\bea
L^u_{eff} & = & -\frac{\lambda^{\prime}_{i12} \lambda^{\prime*}_{i13}}
{2 m_{ \widetilde{e}_i}^2} \bar u_\alpha \gamma_\mu \gamma_L u_\beta
\, \bar s_\beta \gamma_\mu \gamma_R b_\alpha ~, \nn\\
L^c_{eff} & = & -\frac{\lambda^{\prime}_{i22} \lambda^{\prime*}_{i23}}
{2 m_{ \widetilde{e}_i}^2} \bar c_\alpha \gamma_\mu \gamma_L c_\beta
\, \bar s_\beta \gamma_\mu \gamma_R b_\alpha ~.
\eea
For the decays $B \to K \pi$, the term in $ L^u_{eff}$ can contribute
directly or through rescattering of the $u$-quark loop in the BBNS
method. The quantity $r$, which is the ratio of the imaginary part of
the amplitude relative to its real part, is given by \cite{BBNS}
\bea 
r & = & \frac{C_F \alpha_s(m_b)}{4 \pi N_c}Im [{\hat G}_K(s_u)]\nn\\
{\hat G}_K(s) & = & \int^1_0 d x G(s_u,1-x) \Phi_p^K(x) \nn\\ G(s,x) &
= & -4 \int^1_0 u(1-u) ln[s-u(1-u)x] ~,
\eea 
where $C_F=(N_c^2-1)/(2N_c)$, $s_u=(m_u/m_b)^2$ and $\Phi_p^K(x)$ is
the light-cone distribution(LCD) of the kaon . Using the asymptotic
LCD and the fact that $s_u =0$ to a very good approximation, we obtain
\bea 
r & = & \frac{C_F \alpha_s(m_b)}{4 \pi N_c}\frac{2 \pi}{3} \approx
0.015
\label{npphase}
\eea
where we have taken $\alpha_s(m_b)=0.2$. Clearly the NP strong phase
from Eq.~(\ref{npphase}) is $ \sim$ 1$^0$, and is negligible. The
factor of $\alpha_s(m_b)/{(4 \pi) }\approx 0.016$ comes from the
strong coupling constant and the loop factor, and produces a
significant suppression of the rescattering. Only in the SM can this
rescattering be significant, as it is proportional to the large Wilson
coefficient $C_1$. There are also rescattering effects such as the
vertex contributions in the BBNS approach, which will generate new
operators with different colour structure than $ L^u_{eff}$. However
we expect these operators to be suppressed by the factor of
$\alpha_s(m_b)/{(4 \pi) }\approx 0.016$ relative to $ L^u_{eff}$ and
hence to be negligible. The Lagrangian $ L^c_{eff}$ can contribute to
$ K \pi$ final state only through rescattering and will be suppressed
by the same factor $\alpha_s(m_b)/{(4 \pi) }$ relative to $L^u_{eff}$.
It can therefore also be neglected.

For $\btos d{\bar{d}}$, there are four terms in the Lagrangian:
\bea
L^d_{eff}&=&
\frac{\lambda^{\prime}_{i11} \lambda^{\prime*}_{i23}}{  m_{ \widetilde{\nu}_i}^2}
\bar d \gamma_L d \,
\bar {s}\gamma_R b
+\frac{\lambda^{\prime}_{i32} 
\lambda^{\prime*}_{i11}}{  m_{ \widetilde{\nu}_i}^2}
\bar d \gamma_R d \, \bar s \gamma_L b \nonumber \\
&-&\frac{\lambda^{\prime}_{i12} \lambda^{\prime*}_{i13}}
{2  m_{ \widetilde{\nu}_i}^2}
\bar d_\alpha \gamma_\mu \gamma_L d_\beta \, \bar s_\beta \gamma_\mu 
\gamma_R b_\alpha
-\frac{\lambda^{\prime}_{i31} \lambda^{\prime *}_{i21}}
{2  m_{ \widetilde{\nu}_i}^2}
\bar d_\alpha \gamma_\mu \gamma_R d_\beta \, \bar s_\beta \gamma_\mu \gamma_L 
b_\alpha ~.
\eea
For the decays $B \to K \pi$, the terms in $ L^d_{eff}$ can contribute
directly or through rescattering of the $d$-quark loop in the BBNS
method. The terms involving vector operators will have the same
rescatterieng as $L^u_{eff}$ and so the resulting strong phases can
once again be neglected. The scalar operators can only have
electroweak rescattering though the the exchange of a photon or a $Z$
and will therefore be even smaller.

Finally, the relevant Lagrangian for the $\btos s {\bar{s}}$
transition is
\beq
L^s_{eff} = \frac{\lambda^{\prime}_{i32} \lambda^{\prime*}_{i22}} { m_{
\widetilde{\nu}_i}^2} \bar s \gamma_R s \, \bar {s}\gamma_L b+
\frac{\lambda^{\prime}_{i22} \lambda^{\prime*}_{i23}} { m_{
\widetilde{\nu}_i}^2} \bar s \gamma_L s \, \bar {s}\gamma_R b ~.
\eeq
The terms in $L^s_{eff}$ can contribute to the $K \pi$ final state
only through rescattering. They will therefore be suppressed by the
factor $\alpha_s(m_b)/{(4 \pi)}$ relative to $ L^{u,d}_{eff}$, and
hence can be neglected. Note that the operators $ \bar{s}b {\bar b} b$
and $ \bar{s}b {\bar t} t$ can also contribute through rescattering
but will have no imaginary contribution as the gluon momentum is below
the threshold of $b \bar{b}$ and $t \bar{t}$ production.

We have therefore shown in general terms, and within a specific model
of NP, that the NP strong phases are small compared to those of the
SM, using the BBNS approach. As mentioned above, this result is
independent of the specific model of NP, as well as of the method used
to calculate nonleptonic decays.

%%%%%%%%%%%%%%%%%%%%% REFERENCES %%%%%%%%%%%%%%%%%%%%%%%%%%%%%%%%


\begin{thebibliography}{99}

\bibitem{pdg} S.~Eidelman {\it et al.}  [Particle Data Group
Collaboration], Phys.\ Lett.\ B {\bf 592} (2004) 1,
http://pdg.lbl.gov/pdg.html.
%%CITATION = PHLTA,B592,1;%%

\bibitem{CPreview} For a review, see, for example, {\it The BaBar
physics book: Physics at an asymmetric B factory}, eds.\
P.~F.~Harrison and H.~R.~Quinn [BABAR Collaboration], SLAC-R-0504,
October 1998. {\it Papers from Workshop on Physics at an Asymmetric B
Factory (BaBar Collaboration Meeting), Rome, Italy, 11-14 Nov 1996,
Princeton, NJ, 17-20 Mar 1997, Orsay, France, 16-19 Jun 1997 and
Pasadena, CA, 22-24 Sep 1997.}

\bibitem{Sakai} Y. Sakai, talk given at the {\it 32nd International
Conference on High Energy Physics (ICHEP 04)}, Beijing, China, August
2004, \\
http://www.ihep.ac.cn/data/ichep04/ppt/plenary/p11-sakai-y4.pdf

\bibitem{Giorgi} M. Giorgi, talk given at the {\it 32nd International
Conference on High Energy Physics (ICHEP 04)}, Beijing, China, August
2004, \\
http://www.ihep.ac.cn/data/ichep04/ppt/plenary/p12-giorgi-m2.pdf

\bibitem{BKpidecays} Experiment: B.~Aubert {\it et al.}  [BABAR
Collaboration], Phys.\ Rev.\ Lett.\ {\bf 89}, 281802 (2002);
arXiv:hep-ex/0408062, arXiv:hep-ex/0408080, arXiv:hep-ex/0408081;
A.~Bornheim {\it et al.}  [CLEO Collaboration], Phys.\ Rev.\ D {\bf
68}, 052002 (2003); Y.~Chao {\it et al.}  [Belle Collaboration],
Phys.\ Rev.\ D {\bf 69}, 111102 (2004). Theory: A.~J.~Buras,
R.~Fleischer, S.~Recksiegel and F.~Schwab, Eur.\ Phys.\ J.\ C {\bf
32}, 45 (2003), Phys.\ Rev.\ Lett.\ {\bf 92}, 101804 (2004), Nucl.\
Phys.\ B {\bf 697}, 133 (2004), arXiv:hep-ph/0410407; V.~Barger,
C.~W.~Chiang, P.~Langacker and H.~S.~Lee, Phys.\ Lett.\ B {\bf 598},
218 (2004); S.~Mishima and T.~Yoshikawa, arXiv:hep-ph/0408090;
Y.~L.~Wu and Y.~F.~Zhou, arXiv:hep-ph/0409221; H.~Y.~Cheng, C.~K.~Chua
and A.~Soni, arXiv:hep-ph/0409317; Y.~Y.~Charng and H.~n.~Li,
arXiv:hep-ph/0410005; X.~G.~He and B.~H.~J.~McKellar,
arXiv:hep-ph/0410098.
%%CITATION = HEP-EX 0302026;%%
%%CITATION = HEP-EX 0408080;%%
%%CITATION = HEP-EX 0408081;%%
%%CITATION = HEP-EX 0408062;%%
%%CITATION = HEP-EX 0311061;%%
%%CITATION = HEP-EX 0207055;%%
%%CITATION = HEP-PH 0309012;%%
%%CITATION = HEP-PH 0312259;%%
%%CITATION = HEP-PH 0402112;%%
%%CITATION = HEP-PH 0410407;%%
%%CITATION = HEP-PH 0406126;%%
%%CITATION = HEP-PH 0408090;%%
%%CITATION = HEP-PH 0409221;%%
%%CITATION = HEP-PH 0409317;%%
%%CITATION = HEP-PH 0410005;%%
%%CITATION = HEP-PH 0410098;%%

\bibitem{BVVTP} For a study of triple products in the SM and with new
physics, see A.~Datta and D.~London, Int.\ J.\ Mod.\ Phys.\ A {\bf
19}, 2505 (2004).
%%CITATION = HEP-PH 0303159;%%

\bibitem{BaBarTP} B.~Aubert {\it et al.} [BABAR Collaboration],
arXiv:hep-ex/0408017. Note that the earlier Belle measurements of the
same quantities do not show any signs of a nonzero triple-product
asymmetry, see K.-F. Chen {\it et al.}  [Belle Collaboration], Phys.\
Rev.\ Lett.\ {\bf 91}, 201801 (2003).
%%CITATION = HEP-EX 0408017;%%
%%CITATION = HEP-EX 0307014;%%

\bibitem{DLNP} A.~Datta and D.~London, Phys.\ Lett.\ B {\bf 595}, 453
(2004).
%%CITATION = HEP-PH 0404130;%%

\bibitem{PQCD} Y.~Y.~Keum, H.~n.~Li and A.~I.~Sanda, Phys.\ Lett.\ B
{\bf 504}, 6 (2001).
%%CITATION = HEP-PH 0004004;%%

\bibitem{ZFCNC} The model with $Z$-mediated FCNC's was first
introduced in Y.~Nir and D.~J.~Silverman, Phys.\ Rev.\ D {\bf 42},
1477 (1990).
%%CITATION = PHRVA,D42,1477;%%

\bibitem{Z'FCNC} E.~Nardi, Phys.\ Rev.\ D {\bf 48}, 1240 (1993);
J.~Bernabeu, E.~Nardi and D.~Tommasini, Nucl.\ Phys.\ B {\bf 409}, 69
(1993); K.~Leroux and D.~London, Phys.\ Lett.\ B {\bf 526}, 97 (2002);
P.~Langacker and M.~Plumacher, Phys.\ Rev.\ D {\bf 62}, 013006 (2000).
%%CITATION = HEP-PH 9209223;%%
%%CITATION = HEP-PH 9306251;%%
%%CITATION = HEP-PH 0111246;%%
%%CITATION = HEP-PH 0001204;%%

\bibitem{chromo} A. Kagan, Phys.\ Rev.\ D {\bf 51}, 6196 (1995).
%%CITATION = HEP-PH 9409215;%%

\bibitem{LeeGeorgi} T.~D.~Lee, Phys.\ Rev.\ D {\bf 8} (1973) 1226,
Phys.\ Rept.\ {\bf 9} (1974) 143; H.~Georgi, Hadronic J.\ {\bf 1}, 155
(1978).
%%CITATION = PHRVA,D8,1226;%%
%%CITATION = PRPLC,9,143;%%
%%CITATION = HADJM,1,155;%%

\bibitem{IPLL} M.~Imbeault, A.~L.~Lemerle, V.~Page and D.~London,
Phys.\ Rev.\ Lett.\ {\bf 92}, 081801 (2004).
%%CITATION = HEP-PH 0309061;%%

\bibitem{LSSbounds} D.~London, N.~Sinha and R.~Sinha, Europhys.\
Lett.\ {\bf 67}, 579 (2004), Phys.\ Rev.\ D {\bf 69}, 114013 (2004).
%%CITATION = HEP-PH 0304230;%%
%%CITATION = HEP-PH 0402214;%%

\bibitem{Bpenguin} A.~Datta and D.~London, JHEP {\bf 0404}, 072
(2004). See also A.~Datta and D.~London, Phys.\ Lett.\ B {\bf 584}, 81
(2004); J.~Albert, A.~Datta and D.~London, arXiv:hep-ph/0410015.
%%CITATION = HEP-PH 0403165;%%
%%CITATION = HEP-PH 0310252;%%
%%CITATION = HEP-PH 0410015;%%

\bibitem{cquarkconv} M.~Gronau and J.~L.~Rosner, Phys.\ Rev.\ D {\bf
66}, 053003 (2002) [Erratum-ibid.\ D {\bf 66}, 119901 (2002)].
%%CITATION = HEP-PH 0205323;%%

\bibitem{BDK} M. Gronau and D. Wyler, \plb{265}{1991}{172}; D. Atwood,
I. Dunietz and A. Soni, \prl{78}{1997}{3257}. See also M. Gronau and
D. London, \plb{253}{1991}{483}; I. Dunietz, \plb{270}{1991}{75};
N. Sinha and R. Sinha, \prl{80}{1998}{3706}.
%%CITATION = PHLTA,B265,172;%%
%%CITATION = HEP-PH 9612433;%%
%%CITATION = PHLTA,B253,483;%%
%%CITATION = PHLTA,B270,75;%%
%%CITATION = HEP-PH 9712502;%%

\bibitem{Bpipi} M. Gronau and D. London, \prl{65}{1990}{3381}.
%%CITATION = PRLTA,65,3381;%%

\bibitem{Brhopi} A.E. Snyder and H.R. Quinn, \prd{48}{93}{2139};
H.R. Quinn and J.P. Silva, \prd{62}{2000}{054002}.
%%CITATION = PHRVA,D48,2139;%%
%%CITATION = HEP-PH 0001290;%%

\bibitem{Brhorho} L.~Roos [BABAR Collaboration], arXiv:hep-ex/0407051.
%%CITATION = HEP-EX 0407051;%%

\bibitem{pi0refs} Note that this is not the case if the $\pi^0$ comes
from a secondary decay. For example, the BaBar and Belle experiments
have been able to measure the photon polarization in $\bd \to K^*
\gamma$, with $K^* \to \ks\pi^0$, see B.~Aubert {\it et al.}  [BABAR
Collaboration], Phys.\ Rev.\ Lett.\ {\bf 93}, 201801 (2004); K.~Abe
[the Belle Collaboration], arXiv:hep-ex/0411056.
%%CITATION = HEP-EX 0405082;%%
%%CITATION = HEP-EX 0411056;%%

\bibitem{bsKK} R. Fleischer, \plb{459}{1999}{306}.
%%CITATION = HEP-PH 9903456;%%

\bibitem{beneke} M. Beneke, eConf {\bf C0304052}, FO001 (2003)
[arXiv:hep-ph/0308040].
%%CITATION = HEP-PH 0308040;%%

\bibitem{BBNS} M. Beneke, G. Buchalla, M. Neubert and
C.T. Sachrajda, \prl{83}{1999}{1914}, \npb{591}{2000}{313},
\npb{606}{2001}{245}.
%%CITATION = HEP-PH 9905312;%%
%%CITATION = HEP-PH 0006124;%%
%%CITATION = HEP-PH 0104110;%%

\bibitem{charming} M. Ciuchini, E. Franco, G. Martinelli, M. Pierini
and L. Silvestrini, \plb{515}{2001}{33}.
%%CITATION = HEP-PH 0104126;%%

\bibitem{diagrams} M. Gronau, O.F. Hern\' andez, D. London and
J.L. Rosner, \prd{50}{1994}{4529}.
%%CITATION = HEP-PH 9404283;%%

\bibitem{su3} J.P. Silva and L. Wolfenstein, \prd{49}{1994}{1151};
M. Gronau, J.L. Rosner and D. London, \prl{73}{1994}{21};
O.F. Hern\'andez, D. London, M. Gronau and J.L. Rosner,
\plb{333}{1994}{500}; M. Gronau, O.F. Hern\'andez, D. London and
J.L. Rosner, \prd{52}{1995}{6356}, \prd{52}{1995}{6374}.
%%CITATION = HEP-PH 9309283;%%
%%CITATION = HEP-PH 9404282;%%
%%CITATION = HEP-PH 9404281;%%
%%CITATION = HEP-PH 9504326;%%
%%CITATION = HEP-PH 9504327;%%

\bibitem{EWPs} M. Neubert and J.L. Rosner, \plb{441}{1998}{403},
\prl{81}{1998}{5076}; M. Gronau, D. Pirjol and T.M. Yan,
\newprd{60}{1999}{034021} [Erratum-ibid.\ D {\bf 69}, 119901 (2004)].
%%CITATION = HEP-PH 9808493;%%
%%CITATION = HEP-PH 9809311;%%
%%CITATION = HEP-PH 9810482;%%

\bibitem{Charles} For $B\to\pi\pi$ decays, it is more convenient to
use the $t$-quark convention, in which case only a single weak phase
appears: $\alpha$. Then all theoretical parameters can be extracted
from $B \to \pi \pi$ measurements alone, see J.~Charles, Phys.\ Rev.\
D {\bf 59}, 054007 (1999); D.~London, N.~Sinha and R.~Sinha, Phys.\
Rev.\ D {\bf 63}, 054015 (2001).
%%CITATION = HEP-PH 9806468;%%
%%CITATION = HEP-PH 0010174;%%

\bibitem{CGRS} C.~W.~Chiang, M.~Gronau, J.~L.~Rosner and D.~A.~Suprun,
Phys.\ Rev.\ D {\bf 70}, 034020 (2004).
%%CITATION = HEP-PH 0404073;%%

\bibitem{kagan} See, for example, A. Kagan, \plb{601}{2004}{151}.
%%CITATION = HEP-PH 0405134;%%

\bibitem{glambda} N. Sinha and R. Sinha, \prl{80}{1998}{3706};
  A.S. Dighe, I. Dunietz and R. Fleischer, \epjc{6}{1999}{647}.
%%CITATION = HEP-PH 9712502;%%
%%CITATION = HEP-PH 9804253;%%

\bibitem{phiKsNPZFCNC} G.~Hiller, Phys.\ Rev.\ D {\bf 66}, 071502
(2002); A.~K.~Giri and R.~Mohanta, Phys.\ Rev.\ D {\bf 68}, 014020
(2003); D.~Atwood and G.~Hiller, arXiv:hep-ph/0307251; V.~Barger,
C.~W.~Chiang, P.~Langacker and H.~S.~Lee, Ref.~\cite{BKpidecays},
N.~G.~Deshpande and D.~K.~Ghosh, Phys.\ Lett.\ B {\bf 593}, 135
(2004).
%%CITATION = HEP-PH 0207356;%%
%%CITATION = HEP-PH 0306041;%%
%%CITATION = HEP-PH 0307251;%%
%%CITATION = HEP-PH 0310073;%%
%%CITATION = HEP-PH 0311332;%%

\bibitem{phiKsNPRpar} A.~Datta, Phys.\ Rev.\ D {\bf 66}, 071702
(2002); B.~Dutta, C.~S.~Kim and S.~Oh, Phys.\ Rev.\ Lett.\ {\bf 90},
011801 (2003); A.~Kundu and T.~Mitra, Phys.\ Rev.\ D {\bf 67}, 116005
(2003);
%%CITATION = HEP-PH 0208016;%%
%%CITATION = HEP-PH 0208226;%%
%%CITATION = HEP-PH 0302123;%%

\bibitem{phiKsNP} M.~Ciuchini and L.~Silvestrini, Phys.\ Rev.\ Lett.\
{\bf 89}, 231802 (2002); M.~Raidal, Phys.\ Rev.\ Lett.\ {\bf 89},
231803 (2002); J.~P.~Lee and K.~Y.~Lee, Eur.\ Phys.\ J.\ C {\bf 29},
373 (2003); S.~Khalil and E.~Kou, Phys.\ Rev.\ D {\bf 67}, 055009
(2003); M.~Ciuchini, E.~Franco, A.~Masiero and L.~Silvestrini, Phys.\
Rev.\ D {\bf 67}, 075016 (2003) [Erratum-ibid.\ D {\bf 68}, 079901
(2003)]; S.~Baek, Phys.\ Rev.\ D {\bf 67}, 096004 (2003); C.~W.~Chiang
and J.~L.~Rosner, Phys.\ Rev.\ D {\bf 68}, 014007 (2003); K.~Agashe
and C.~D.~Carone, Phys.\ Rev.\ D {\bf 68}, 035017 (2003); G.~L.~Kane,
P.~Ko, H.~b.~Wang, C.~Kolda, J.~h.~Park and L.~T.~Wang, Phys.\ Rev.\
Lett.\ {\bf 90}, 141803 (2003); R.~Arnowitt, B.~Dutta and B.~Hu,
Phys.\ Rev.\ D {\bf 68}, 075008 (2003); C.~S.~Huang and S.~h.~Zhu,
Phys.\ Rev.\ D {\bf 68}, 114020 (2003); M.~Frank, Phys.\ Rev.\ D {\bf
68}, 035011 (2003); J.~F.~Cheng, C.~S.~Huang and X.~h.~Wu, Phys.\
Lett.\ B {\bf 585}, 287 (2004), Nucl.\ Phys.\ B {\bf 701}, 54 (2004);
C.~Dariescu, M.~A.~Dariescu, N.~G.~Deshpande and D.~K.~Ghosh, Phys.\
Rev.\ D {\bf 69}, 112003 (2004); P.~Ball, S.~Khalil and E.~Kou, Phys.\
Rev.\ D {\bf 69}, 115011 (2004); Y.~L.~Wu and Y.~F.~Zhou, Eur.\ Phys.\
J.\ C {\bf 36}, 89 (2004).
%%CITATION = HEP-PH 0208087;%%
%%CITATION = HEP-PH 0208091;%%
%%CITATION = HEP-PH 0209290;%%
%%CITATION = HEP-PH 0212397;%%
%%CITATION = HEP-PH 0212023;%%
%%CITATION = HEP-PH 0301269;%%
%%CITATION = HEP-PH 0302094;%%
%%CITATION = HEP-PH 0304229;%%
%%CITATION = HEP-PH 0304239;%%
%%CITATION = HEP-PH 0307152;%%
%%CITATION = HEP-PH 0307354;%%
%%CITATION = PHRVA,D68,035011;%%
%%CITATION = HEP-PH 0306086;%%
%%CITATION = HEP-PH 0308305;%%
%%CITATION = HEP-PH 0311361;%%
%%CITATION = HEP-PH 0403252;%%
%%CITATION = HEP-PH 0404055;%%

\bibitem{Proton} I. Hinchliffe and T. Kaeding, \prd{47}{1993}{279};
C.E. Carlson, P. Roy and M. Sher, \plb{357}{1995}{99}; A.Y. Smirnov
and F. Vissani, \plb{380}{1996}{317}.
%%CITATION = PHRVA,D47,279;%%
%%CITATION = HEP-PH 9506328;%%
%%CITATION = HEP-PH 9601387;%%

\bibitem{DatXin} See for example A. Datta, J.M. Yang, B.L. Young and
X. Zhang, \prd{56}{1997}{3107}.
%%CITATION = HEP-PH 9704257;%%

\bibitem{lambdabNP} W. Bensalem, A. Datta and D. London,
\prd{66}{2002}{094004}.
%%CITATION = HEP-PH 0208054;%%

\bibitem{BuraseffH} See, for example, G. Buchalla, A.J. Buras and
  M.E. Lautenbacher, {\it Rev.\ Mod.\ Phys.} {\bf 68}, 1125 (1996),
  A.J. Buras, ``Weak Hamiltonian, CP Violation and Rare Decays,'' in
  {\it Probing the Standard Model of Particle Interactions}, ed.\
  F. David and R. Gupta (Elsevier Science B.V., 1998), pp.\ 281-539.
%%CITATION = HEP-PH 9512380;%%
%%CITATION = HEP-PH 9806471;%%.

\end{thebibliography}
\end{document}